\documentclass[useAMS,usenatbib,usegraphicx]{mn2e}

\newcommand{\singlecolsize}{0.475}
\newcommand{\doublecolsize}{1.0}

\newcommand{\sqdeg}{deg$^{2}$}
\newcommand{\persqdeg}{deg$^{-2}$}
\newcommand{\cosmos}{$(\Omega_{m},\Omega_{\Lambda})_0$}
\newcommand{\Vmax}{$V_{\rm max}$}
\newcommand{\phistar}{$\phi^{*}$}
\newcommand{\mstar}{$M^{*}$}
\newcommand{\uband}{$^{0}u$}
\newcommand{\usband}{$^{0.1}u$}

\newcommand{\dd}{{\rm d}}
\newcommand{\sgsep}{\Delta_{\rm sg}}
\newcommand{\rsb}{\mu_{r,50}}
\newcommand{\upetro}{u_{\rm petro}}
\newcommand{\gpetro}{g_{\rm petro}}
\newcommand{\rpetro}{r_{\rm petro}}
\newcommand{\ipetro}{i_{\rm petro}}
\newcommand{\umodel}{u_{\rm model}}

\newcommand{\rmodel}{r_{\rm model}}

\newcommand{\upsf}{u_{\rm psf}}
\newcommand{\gpsf}{g_{\rm psf}}
\newcommand{\rpsf}{r_{\rm psf}}
\newcommand{\ipsf}{i_{\rm psf}}

\newcommand{\jddate}{JD 24\,53228}

\title[SDSS $u$-band Galaxy Survey]{The SDSS $u$-band Galaxy Survey: 
  Luminosity functions and evolution}
\author[I.~K.~Baldry et al.]{I.~K.~Baldry$^1$, K.~Glazebrook$^1$,
  T.~Budav\'ari$^1$, D.~J.~Eisenstein$^2$,\newauthor
  J.~Annis$^3$, N.~A.~Bahcall$^4$, 
  M.~R.~Blanton$^5$, J.~Brinkmann$^6$, I.~Csabai$^{1,7}$,\newauthor
  T.~M.~Heckman$^1$, H.~Lin$^3$, 
  J.~Loveday$^8$, R.~C.~Nichol$^9$, D.~P.~Schneider$^{10}$\\
$^1${Department of Physics and Astronomy, Johns Hopkins University,
3400 North Charles Street, Baltimore, MD~21218, USA.\label{JHU}}\\
$^2${Steward Observatory, 933 North Cherry Avenue, Tucson, 
AZ~85721, USA.\label{Arizona}}\\
$^3${Fermi National Accelerator Laboratory, P.O.\ Box 500, 
Batavia, IL~60510, USA.\label{Fermilab}}\\
$^4${Department of Astrophysical Sciences, Princeton University, 
Princeton, NJ~08544, USA.\label{Princeton}}\\
$^5${Center for Cosmology and Particle Physics,
Department of Physics, New York University, 4 Washington Place, 
NY~10003, USA.\label{NYU}}\\
$^6${Apache Point Observatory, P.O. Box 59, Sunspot, 
NM 88349, USA.\label{APO}}\\
$^7${Department of Physics of Complex Systems, 
E\"{o}tv\"{o}s Lor\'and University, Pf.~32, H-1518 Budapest, 
Hungary.\label{Eotvos}}\\
$^8${Astronomy Centre, University of Sussex, Falmer, 
Brighton, BN1~9QJ, UK.\label{Sussex}}\\
$^9${Institute of Cosmology and Gravitation, Mercantile House, 
Hampshire Terrace, University of Portsmouth, PO1~2EG, UK.\label{Portsmouth}}\\
$^{10}${Department of Astronomy and Astrophysics, 
525 Davey Laboratory, Pennsylvania State University, 
PA~16802, USA.\label{PSU}}
}

\begin{document}

\date{Submitted 2004 November 26; accepted 2005 January 7.}

\pagerange{\pageref{firstpage}--\pageref{lastpage}} \pubyear{2005}

\maketitle

\label{firstpage}

\begin{abstract}
  We construct and analyze a $u$-band selected galaxy sample from the SDSS
  Southern Survey, which covers 275\,\sqdeg. The sample includes 43223
  galaxies with spectroscopic redshifts in the range $0.005<z<0.3$ and with
  $14.5<u<20.5$. The S/N in the $u$-band Petrosian aperture is improved by
  coadding multiple epochs of imaging data and by including sky-subtraction
  corrections.  Luminosity functions for the near-UV \usband\ band
  ($\lambda\approx322\pm26{\rm\,nm}$) are determined in redshift slices of
  width 0.02, which show a highly significant evolution in \mstar\ of
  $-0.8\pm0.1$ mag between $z=0$ and 0.3; with
  $M^{*}-5\log h_{70}=-18.84\pm0.05$ (AB mag), 
  $\log\phi^{*}=-2.06\pm0.03$ ($h_{70}^3{\rm\:Mpc^{-3}}$) and 
  $\log\rho_L=19.11\pm0.02$ ($h_{70}{\rm\:W\,Hz^{-1}\,Mpc^{-3}}$) at $z=0.1$.
  The faint-end slope determined for $z<0.06$ is given by
  $\alpha=-1.05\pm0.08$.  This is in agreement with recent determinations from
  GALEX at shorter wavelengths.  Comparing our $z<0.3$ luminosity density
  measurements with $0.2<z<1.2$ from COMBO-17, we find that the 280-nm density
  evolves as $\rho_L\propto(1+z)^{\beta}$ with $\beta=2.1\pm0.2$; and find no
  evidence for any change in slope over this redshift range.  By comparing
  with other measurements of cosmic star formation history, we estimate that
  the effective dust attenuation at 280\,nm has increased by $0.8\pm0.3$ mag
  between $z=0$ and 1.
\end{abstract}


\section{Introduction}
\label{sec:intro}

In the absence of dust, the rest-frame UV luminosity of a galaxy is nearly
proportional to the total mass of short-lived OB stars and therefore to the
star formation rate (SFR).  This has been used to show that the
volume-averaged SFR of the Universe has been declining since at least $z\sim1$
\citep{lilly96,madau96,MPD98,CSB99,steidel99,wilson02}.  Until the
observations of the Galaxy Evolution Explorer (GALEX) \citep{martin05}, the
accuracy in the measured evolution rate had been low because of the lack of
100--300\,nm surveys at low redshift. The FOCA ballon-borne telescope
\citep{milliard92} had been the only instrument for measuring the galaxy
luminosity function (LF) at these wavelengths; this survey covered about
2\,\sqdeg\ \citep{treyer98,sullivan00}. However, `sun-tanning bands'
$u$/$U$/$U'$ also provide a window on the star-forming properties of galaxies
as they are significantly more sensitive to young stellar populations than the
$B$/$b_J$/$g$ bands; over 5000\,\sqdeg\ has been imaged in the $u$-band as
part of the Sloan Digital Sky Survey (SDSS).  Using these data,
\citet{hopkins03} have demonstrated that the $u$-band luminosity is a
reasonable measure of the SFR by comparing with other SFR indicators
(H$\alpha$, O{\small II} and far-IR).

\begin{figure}
\includegraphics[width=\singlecolsize\textwidth]{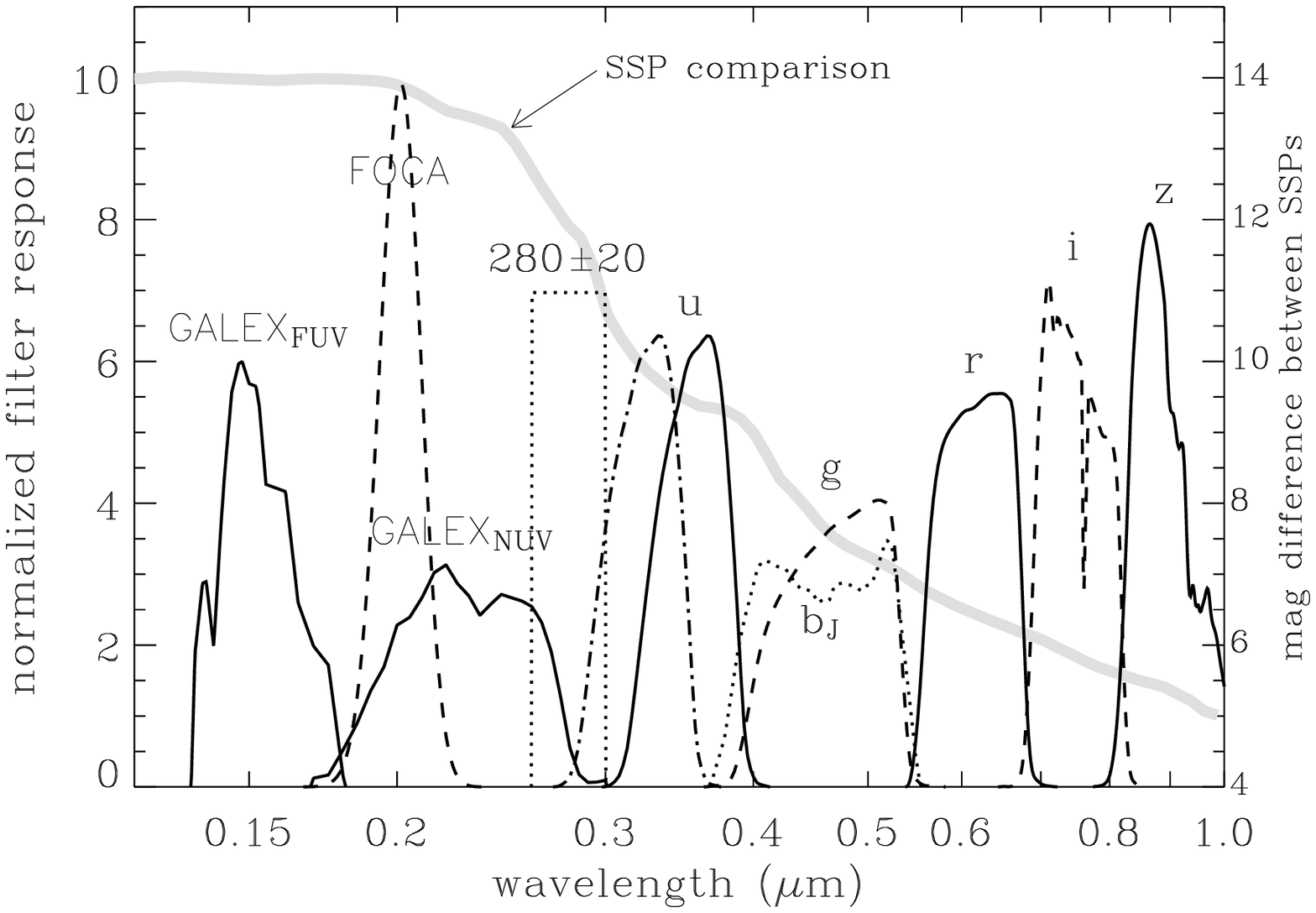}
\caption{Filter profiles: GALEX \citep{martin03}; FOCA \citep{milliard92};
  SDSS $ugriz$ \citep{stoughton02}; APM $b_J$; and an artificial 280-nm band
  \citep{wolf03}. The $u$-band shifted to $z=0.1$ is shown by the
  dash-and-dotted line (\usband\ band). Each curve is normalized so that the
  integral of the normalized transmission $T\,\dd\ln\lambda$ is equal to
  unity.  Thus the height of each curve is related to its resolving power
  ($\lambda/\Delta\lambda$). The thick gray line represents the difference in
  magnitudes between dust-free 5\,Myr and 10000\,Myr simple stellar
  populations from the models of \citet{BC03} (the $y$-axis for this line is
  shown on the right).}
\label{fig:filters}
\end{figure}

Figure~\ref{fig:filters} shows the normalized responses of the GALEX, FOCA and
SDSS filters. The difference between the spectra of a young and an old stellar
population is also plotted. This shows (i) the importance of the $u$-band in
constraining the spectral energy distributions of galaxies between the GALEX
UV bands and the visible bands, and (ii) the significantly increased
sensitivity to young stellar populations of the $u$-band in comparison with
the $g$-band.

While imaging data can be used to study correlations between properties of
galaxies, in order to measure the space density of sources, it is preferable
to select the targets for spectroscopic followup using the most appropriate
band.  The largest spectroscopic surveys at $z<0.2$ are the SDSS
\citep{york00}, the 2dF Galaxy Redshift Survey (2dFGRS) \citep{colless01} and
the 6dF Galaxy Survey (6dFGS) \citep{jones04}, which have principally selected
galaxies using the $r$, $b_J$ and $K$ bands, respectively ($\lambda_{\rm
  eff}\approx616$, 456 and 2160\,nm).  While the $b_J$ selection is more
biased toward star-forming galaxies than the other two, the band does not
sample a majority of the light below the rest-frame 400-nm break at $z<0.15$.
In this paper, we describe the SDSS $u$-band Galaxy Survey ($u$GS), which
provides an intermediate selection between the $r$/$b_J$ and the GALEX bands,
and a local sample for comparison with higher redshift surveys that are
selected at similar rest-frame wavelengths (e.g.\ DEEP2, \citealt{davis03}).
The $u$-band integrates flux almost entirely from below the 400-nm break
($\lambda_{\rm eff}\approx355$\,nm).\footnote{The SDSS $u$-band has a small
  amount of contamination from around 710\,nm. This corresponds to about a
  0.02 mag effect for mid-K stars and galaxies of similar color
  \citep{abazajian04}.  We have not corrected for this because most of the
  galaxies in our sample are significantly bluer.}

The plan of the paper is as follows: in \S~\ref{sec:data} we describe the
basics of the SDSS (which can be skipped for those familiar with the main
survey); in \S~\ref{sec:southern-survey} we introduce the Southern Survey; in
\S~\ref{sec:lf} we present the results for the LFs (with more details in
Appendix~\ref{sec:lf-more}); in \S\S~\ref{sec:discuss} \&~\ref{sec:conclude}
we discuss and conclude; and in Appendix~\ref{sec:sky-sub-corr} we outline
sky-subtraction corrections for SDSS $u$-band magnitudes. Note that magnitudes
used in this paper are corrected for Milky-Way (MW) extinction unless
otherwise noted.

\section{The Sloan Digital Sky Survey}
\label{sec:data}

The Sloan Digital Sky Survey is a project, with a dedicated 2.5-m telescope,
designed to image $10^4$\,\sqdeg\ and obtain spectra of $10^6$ objects
\citep{york00,stoughton02,abazajian03,abazajian04}. The imaging covers five
broadbands, $ugriz$ with effective wavelengths of 355, 467, 616, 747 and
892\,nm, using a mosaic CCD camera \citep{gunn98}.  Observations with a 0.5-m
photometric telescope \citep{hogg01} are used to calibrate the 2.5-m telescope
images using a standard star system \citep{fukugita96,smith02}.  Spectra are
obtained using a 640-fiber fed spectrograph with a wavelength range of 380 to
920\,nm and a resolution of $\lambda/\Delta\lambda\sim1800$ \citep{uomoto99}.

The imaging data are astrometrically calibrated \citep{pier03} and the images
are reduced using a pipeline {\tt photo} that measures the observing
conditions, and detects and measures objects. In particular, {\tt photo}
produces various types of magnitude measurement including: (i) `Petrosian
magnitudes', the summed flux in an aperture that depends on the
surface-brightness (SB) profile of the object, a modified version of the flux
quantity defined by \citet{Petrosian76}; (ii) `model magnitudes', a fit to the
flux using the best fit of a \citet{deVaucouleurs59} and an exponential
profile \citep{Freeman70}; (iii) `PSF magnitudes', a fit using the local
point-spread function.  The magnitudes are extinction-corrected using the MW
dust maps of \citet*{SFD98}.  Details of the imaging pipelines are given by
\citet{lupton01} and \citet{stoughton02}.

Once a sufficiently large area of sky has been imaged, the data are analyzed
using `targeting' software routines that determine the objects to be observed
spectroscopically. The targets that are part of the `main program' include:
galaxies with $\rpetro<17.8$ (`MAIN selection'; \citealt{strauss02}); quasars
selected by various color criteria with $\ipsf<19.1$ or 20.2 (`QSO selection';
\citealt{richards02}); and luminous-red galaxies selected by color cuts with
$\rpetro<19.2$ or 19.5 (`LRG selection'; \citealt{eisenstein01}).  The targets
from all the samples are assigned to plates, each with 640 fibers, using a
tiling algorithm \citep{blanton03tile}.

\section{The Southern Survey and Sample Selection}
\label{sec:southern-survey}

The main program of the SDSS is concentrated in the Northern Galactic Pole
(NGP), with only three `stripes' (2.5\degr\ wide) in the Southern Galactic
Pole (SGP).\footnote{In the nomenclature of SDSS, a `stripe' consists of a
  Northern and a Southern `strip' because there are gaps between the
  detectors, which are aligned in six camera columns. The two strips are
  interleaved to produce a contiguously imaged stripe. A `run' is a continuous
  drift-scan observation of a single strip.  See \citet{stoughton02} for
  details on nomenclature including the flags used to select the
  objects.\label{ftn:stripe-strip}} During the times when it is not possible
to observe the NGP, the `Southern Survey' has been in operation. This has
involved repeat imaging of the middle SGP stripe (on the celestial equator)
and additional non-standard spectroscopic observations.  Here we define the
Southern Survey region as RA from $-50.8$\degr\ to 58.6\degr\ (20.6\,h to
3.9\,h) and DEC from $-1.26$\degr\ to 1.26\degr\ (an area of 275.7\,\sqdeg).
Over this region, there are 6 to 18 repeat images depending on the sky
position with 253 unique spectroscopic plates observed (including 57 for the
main program); up to \jddate. The SDSS $u$GS consists of high 
signal-to-noise ratio (S/N) magnitude
measurements using a coadded catalog (\S~\ref{sec:coadd-imaging}),
sky-subtraction corrections to $u$-band Petrosian magnitudes
(Appendix~\ref{sec:sky-sub-corr}), and spectra selected in a variety of ways
(\S~\ref{sec:spec-target}) with completeness corrections for $\upetro<20.5$
(\S~\ref{sec:completeness}). These sections
(\S\S~\ref{sec:coadd-imaging}--\ref{sec:completeness};
Table~\ref{tab:spec-select}; Figs.~\ref{fig:z-histos}--\ref{fig:comp-maps})
can be skipped or browsed for those not interested in the survey details at
this stage.

\subsection{Creating a coadded imaging catalog}
\label{sec:coadd-imaging}

The repeat imaging data can be coadded at the image level or at the catalog
level. The former is necessary for increasing the depth of the imaging.  Here
we are mostly interested in improving the $u$-band Petrosian S/N and
associated colors. Therefore, it is adequate to coadd the data at the catalog
level because even the bluest galaxies in our sample are detected in the $g$
and $r$ bands with adequate S/N to define a Petrosian aperture. This has the
advantage that the data passes through the standard imaging pipeline with no
adjustments.

Our procedure for coadding the catalog was in two parts: (1) producing a
master catalog of groups of matched objects; and (2) selecting appropriate
samples for analysis.

\begin{enumerate}
\renewcommand{\theenumi}{(\arabic{enumi})}

\item First of all, each camera column and `strip' is considered
  separately (see footnote~\ref{ftn:stripe-strip}). Catalog objects
  are selected from all the runs so that they are: (i) unique to a run
  (`status\_ok\_run'); (ii) not near the edge of a frame (not
  `object\_edge') unless they are deblended from an edge object
  (`object\_deblended\_at\_edge'); (iii) not `bright', which refers to
  a preliminary identification of bright objects in the catalog; (iv)
  not `blended' unless they are the final product of a deblending
  process (`object\_nodeblend' or `nchild'=0); and (v) detected in two
  or more of the the five bands, which is to remove cosmic rays and
  other artifacts only detected in one band. All the qualifying
  objects are then matched to objects in other runs within a radius of
  1.5 arcsec.  If there is more than one object from one run in a
  single matched group (which is rare but can occur because of
  deblending), then only the object that has the smallest deviation
  from the median $\rpetro$ value of the group is selected.  Various
  mean and median values are calculated for the matched groups of
  objects, while the object flags are taken from the highest quality
  run.  The SDSS asinh magnitudes \citep*{LGS99} are converted to
  linear units before averaging (Eq.~\ref{eqn:asinh-convert}).  This
  procedure produced a coadded catalog of $6\times10^6$ objects.

\item In addition to selecting on magnitudes, we apply the following criteria:
  (i) the object is detected in at least two bands out of $ugr$ based on the
  best imaging run; (ii) the matched object is from a combination including
  over half of the maximum number of runs available at that sky
  position;\footnote{Complex objects could have their centres shifted by more
    than the matching radius between runs and matching in more than half the
    runs ensures that only one copy of a complex object is included.} (iii)
  the object is not saturated in the fiducial $r$-band; (iv) the S/N of the
  Petrosian flux is greater than three in each of the $u$, $g$ and $r$ bands
  (the uncertainty is determined from the standard error over the coadded
  runs). The $u$-band Petrosian flux is adjusted for sky-subtraction errors
  (Appendix~\ref{sec:sky-sub-corr}) and if there are large differences between
  the median and mean $ugr$ fluxes, the median is used to replace the mean
  value (and a modified standard error that excludes the minimum and maximum
  values is used). For the SDSS $u$GS, we selected resolved sources with
  $\upetro<20.5$.

\end{enumerate}

The data from the coadded imaging catalog that we use for the science results
in this paper are: the mean Petrosian fluxes (median for $\sim3$\% of
objects); the standard errors of the fluxes\footnote{The standard error is
  $\sigma/\sqrt{N}$; while the modified error is given by $1.3\,\sigma_{\rm
    mod}/\sqrt{N-2}$ when using the median flux, where $\sigma_{\rm mod}$ is
  the standard deviation excluding the minimum and maximum values.  The factor
  of 1.3 is determined from the median of $\sigma/\sigma_{\rm mod}$ for the
  data.} ($+2$\% error added in quadrature for $k$-correction fitting); the
median $\rpsf - \rmodel$ values (for star-galaxy separation); the mean sky
positions; and the mean Petrosian half-light radii.  The median S/N of the
coadded-catalog Petrosian $u$-band fluxes at $u\sim20.5$ is 10, which is a
factor of $\sim3$ improvement over a single epoch of imaging.

\subsection{Spectroscopic target selections}
\label{sec:spec-target}

In addition to repeat imaging along the Southern Survey equatorial stripe,
there are also extra spectroscopic observations. There are six general
programs that contribute most of the redshifts to a $\upetro<20.5$ galaxy
sample including specifically designed selection criteria for this survey.

For the selection, star-galaxy separation and SB parameters are defined as
follows:
\begin{eqnarray}
\sgsep & = & \rpsf - \rmodel \label{eqn:sgsep} \\
\rsb   & = & \rpetro + 2.5 \log( 2 \pi R_{r,50}^2 ) \label{eqn:rsb}
\end{eqnarray}
where $R_{r,50}$ is the radius containing half the Petrosian flux.  Thus
$\rsb$ is the mean SB within the Petrosian half-light radius.  Note that most
of the targets were selected using a single epoch of imaging before any
coadded imaging catalog was produced.

\begin{table*}
\caption{Number of objects and spectra for $u$-band selected samples}
\label{tab:spec-select}
\begin{center}
\begin{tabular}{lrrrrrrrrrr} \hline
sample selection & no.~of & 
\multicolumn{8}{c}{no.~spectroscopically observed by SDSS
(broken down by program)\rlap{$^a$}} 
& 2dF\rlap{$^b$} \\
& objects & \ref{itm:main-select}MAIN & \ref{itm:qso-select}QSO 
& \ref{itm:u-select}$u$-band & \ref{itm:lowz-select}low-$z$
& \ref{itm:lowz-pre-select}low-$z$ & \ref{itm:highz-select}high-$z$ 
& other & total & \\
& & & & & & & & & \\
$u<20.0$ (incl.\ unresolved){$^c$}
&321768& 18829& 15061&  5606&  2888&  1027&   139&  8317& 51867&    79\\
$u<20.5$ \& $\sgsep>0.05$ {$^d$}
& 74901& 22141&  6825&  7400&  6687&  1846&   637&   303& 45839&   104\\
$u<21.0$ \& $\sgsep>0.05$ {$^e$}
&146488& 23502& 10089&  7536&  9997&  2200&  2084&   584& 55992&   109\\
\hline
\end{tabular}
\end{center}
\begin{flushleft}
$^a$Number observed by various Southern Survey programs described in
\S~\ref{sec:spec-target} and the total number for all the programs; 
up to \jddate.\newline 
$^b$Number of 2dFGRS redshifts not observed by SDSS. Only two 2dFGRS
fields overlap with the Southern Survey.\newline 
$^c$Sample used to assess star-galaxy separation; see
Fig.~\ref{fig:sgsep-z} for redshift versus $\sgsep$ 
(Eq.~\ref{eqn:sgsep}).\newline 
$^d$Sample used to compute the galaxy LFs (\S~\ref{sec:lf}); see
Fig.~\ref{fig:z-histos} for the redshift histograms,
Fig.~\ref{fig:color-color} for color-color plots, and Fig.~\ref{fig:comp-maps}
for spectroscopic completeness as a function of color.\newline 
$^e$Sample used to assess completeness to fainter magnitudes; see
Fig.~\ref{fig:completeness} for completeness as a function of magnitude.
\end{flushleft}
\end{table*}

\begin{enumerate}
\renewcommand{\theenumi}{(\arabic{enumi})}

\item \label{itm:main-select} {\em MAIN selection}.  Many of the spectra were
  targeted as part of the main galaxy sample. This has the following basic
  criteria:
  \begin{eqnarray}
  \rpetro & < & 17.8 \\
  \rsb    & < & 24.5 \\
  \sgsep  & > & 0.25 \: .
  \end{eqnarray}
  See \citet{strauss02} for details.  In addition to the main program, galaxy
  targets that were missed, because of fiber-placement restrictions or
  photometric errors, were included on additional plates. Thus the
  completeness of the MAIN selection is very high ($\approx99$\%) over much of
  the Southern Survey. 

\item \label{itm:qso-select} {\em QSO selection}. SDSS quasars were selected
  by looking for non-stellar colors using PSF magnitudes. For the low redshift
  candidates there was no requirement that the object be a point source. This
  enables selection of resolved galaxies where the central light may be
  dominated by a quasar but the total light may not be. The low redshift
  (i.e.\ extended sources) selection in the main program can be approximated
  by
  \begin{eqnarray}
  \ipsf   & < & 19.1 \label{eqn:qso1} \\
  \upsf - \gpsf & < & 0.9 \label{eqn:qso2} \\
  \upsf - 0.5\gpsf - 0.5\rpsf & < & 0.9 \label{eqn:qso3} \: .
  \end{eqnarray}
  See \citet{richards02} for details (e.g.\ fig.~13 of that paper). In
  addition to the main program, several alternative target selections were
  made on additional plates. These included probing closer to the stellar
  locus (and also closer to the galaxy locus), in the stellar locus
  \citep{vandenberk05} and selecting fainter QSO candidates.

\item \label{itm:u-select} {\em $u$-band galaxy selection}. Here the idea was
  to was to obtain a complete $u$-band magnitude-limited galaxy sample by
  filling in the redshifts missed by other selections. The criteria were
  mostly as follows:
  \begin{eqnarray}
  u_{\rm select} & < & 19.8 \\
  \gpetro & < & 20.5 \\
  \rpetro & < & 20.5 \\
  \rsb    & < & 24.5 \\
  \sgsep  & > & 0.2 \: .
  \end{eqnarray}
  where $u_{\rm select} = \umodel - \rmodel + \rpetro$, which can be regarded
  as a pseudo-Petrosian $u$-band magnitude. The reason for using this
  magnitude definition was to avoid $\upetro$ (single epoch), which has larger
  Poisson noise and systematic errors (Appendix~\ref{sec:sky-sub-corr}), and
  has a greater susceptibility to imaging artifacts. The $g$ and $r$-band
  requirements were also used to reduce artifacts. On some plates, the
  magnitude criteria were relaxed by 0.2 mag to $u_{\rm select}<20.0$.

\item \label{itm:lowz-select} {\em Low-$z$ galaxy selection}. Targets were
  selected using photometric redshifts with primarily:
  \begin{eqnarray}
  z_{\rm photo} & < & 0.15 \\
  \rpetro & < & 19.0 \\
  \sgsep  & > & 0.15 \: .
  \end{eqnarray}
  The photometric technique was calibrated using spectroscopically confirmed
  redshifts from the main program and the Southern Survey.  In order to test
  selection effects, other photometric redshift ranges were also sparsely
  sampled as were galaxies with $19.0<\rpetro<19.5$.

\item \label{itm:lowz-pre-select} {\em Low-$z$ galaxy selection} (precursor).
  Targets were selected using photometric redshifts with primarily:
  \begin{eqnarray}
  z_{\rm photo} & < & 0.2 \\
  \ipetro & < & 20.0 \\
  \sgsep  & > & 0.15 \: .
  \end{eqnarray}
  These plates were also designed to be complimentary to the higher
  photometric redshift selection described below.

\item \label{itm:highz-select} {\em High-$z$ galaxy selection}. Here the idea
  was to obtain spectroscopic redshifts for non-LRG galaxies above redshift
  0.3 in order to improve photometric redshifts for these types of galaxies.
  The targets were selected with:
  \begin{eqnarray}
  z_{\rm photo} & > & 0.3 \\
  \rmodel & < & 19.5 \\
  \sgsep  & > & 0.15 \: .
  \end{eqnarray}
  In detail, the photometric redshift cut was converted to a series of color
  and magnitude cuts.

\end{enumerate}

\begin{figure*}
\includegraphics[width=0.9\textwidth]{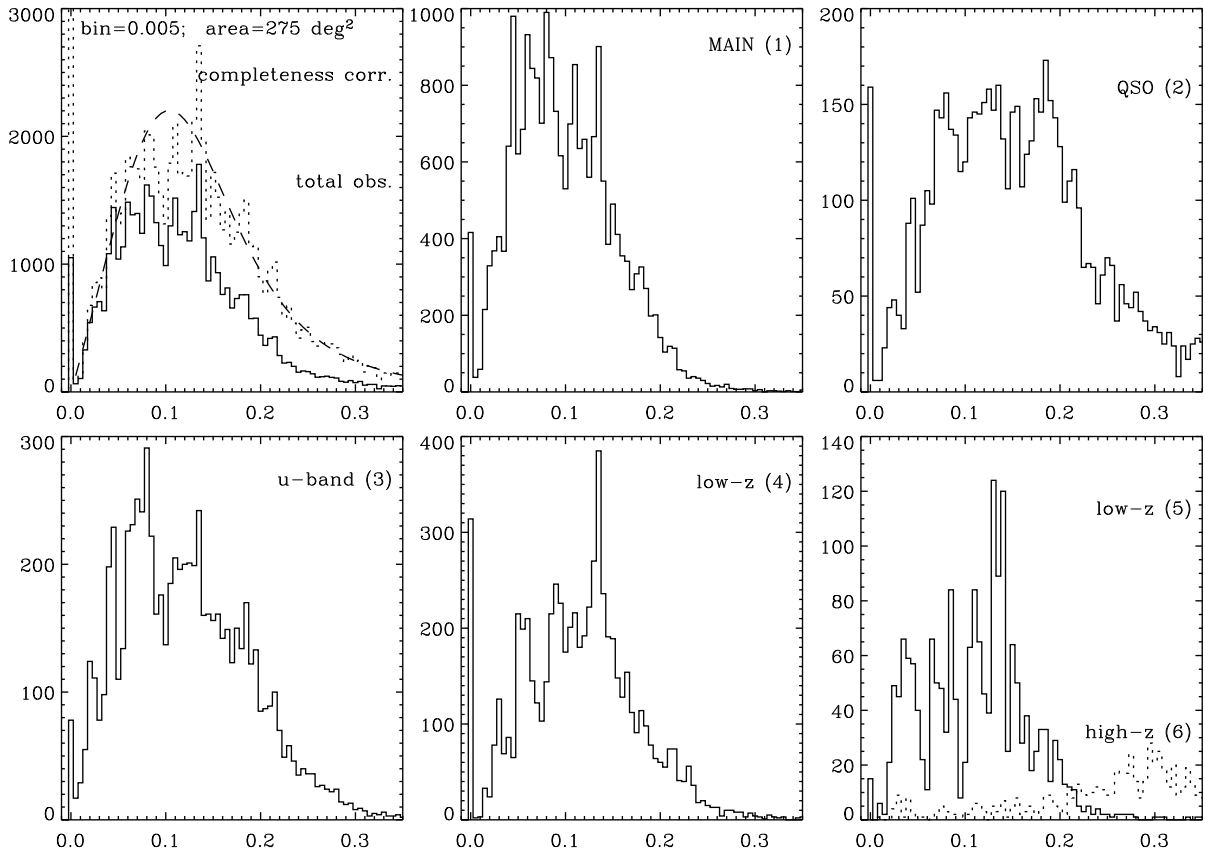}
\caption{Redshift histograms ($\upetro<20.5$ sample): 
  number per 0.005 bin versus redshift.  The top-left panel shows data from
  all the spectroscopic programs; with the solid line representing the
  observed data and the dotted line representing the completeness-corrected
  data (\S~\ref{sec:completeness}). The dashed line is a fit using Eq.~1 of
  \citet{percival01} with parameter values 0.158, 2.054 and 0.603. The other
  five panels show the data from the six programs described in
  \S~\ref{sec:spec-target}. Note the $y$-axis scales vary; and the program
  names refer to the selection algorithms not the spectral classifications.}
\label{fig:z-histos}
\end{figure*}

Spectra were matched to photometric objects within 1.5 arcsec or within
$R_{r,50}$ (1.5--12 arcsec, which accounts for 44 objects where an
earlier version of {\tt photo} may have targeted a deblended piece of the
galaxy).  The number of spectra contributing to various $u$-band selected
samples are shown in Table~\ref{tab:spec-select}.  Figure~\ref{fig:z-histos}
shows the redshift histograms.  Other selections include stellar programs for
unresolved sources and about 100 redshifts were included from the 2dFGRS
(which is similar to $g\la19$ galaxy selection).

\subsection{Completeness corrections}
\label{sec:completeness}

To determine spectroscopic completeness corrections, we do not attempt to back
track and reproduce the above selection criteria. Many of the criteria are
complicated and somewhat arbitrary for creating a $u$-band magnitude-limited
sample. In addition, the photometric code and imaging run varies between
selections and our aim is to select using Petrosian magnitudes from a coadded
catalog. Instead we use an empirical approach for estimating the completeness
factors ($C$), which we define as the fraction of photometric objects that
have been observed spectroscopically.

\begin{figure*}
\includegraphics[width=0.7\textwidth]{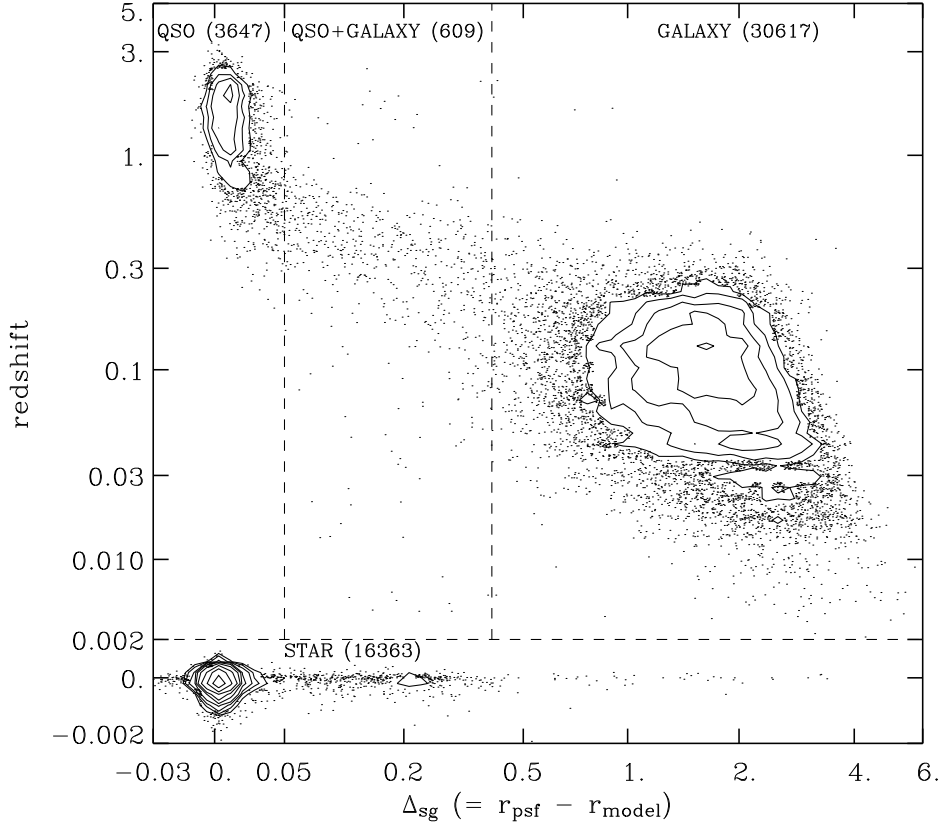}
\caption{Redshift versus star-galaxy separation parameter
  for 51236 objects with $\upetro<20.0$.  The dashed lines divide regions
  where the spectral classification is highly uniform ($>99.9$\% STAR;
  $>99.9$\% QSO; 99\% GALAXY) except for the QSO+GALAXY region, which has a
  mixed classification (77\% and 23\%, respectively). The axes are linear in
  $\log(z+0.004)$ and $\log(\sgsep+0.1)$.  The solid lines represent
  logarithmically-spaced density contours with 4 contours per factor of 10.
  Note that there are no point sources ($\sgsep<0.05$) with redshifts between
  0.002 and 0.195. There are two 1D-spectral routines used for redshift and
  spectral classification within the SDSS collaboration (M.~Subbarao et al.;
  D.~Schlegel et al.\ in preparation). Only redshifts where the two routines
  produced similar results were included in this figure. This rejects only 1\%
  of the measured redshifts.}
\label{fig:sgsep-z}
\end{figure*}

Before describing the completeness corrections, we discuss
star(-quasar)-galaxy separation. For each object we used the median $\sgsep$
value ($=\rpsf-\rmodel$) from the coadded imaging for robustness.
Figure~\ref{fig:sgsep-z} shows redshift versus $\sgsep$ for a spectroscopic
sample with $\upetro<20.0$ (Table~\ref{tab:spec-select}). We define four
regimes: sources with intra-galactic redshifts ($z<0.002$); and unresolved
($\sgsep<0.05$), weakly resolved ($0.05<\sgsep<0.4$) and strongly resolved
($\sgsep>0.4$) sources with extra-galactic redshifts ($z>0.002$).  The main
point to note is that there are no unresolved extra-galactic sources with
redshifts less than about 0.2 in this sample. While there are certainly
selection effects against targeting this type of object, we note that about
7000 of the point sources with redshifts were targeted at random (as part of
testing the QSO selection; \citealt{vandenberk05}).  There are some weakly
resolved sources at low redshift. In this regime, some objects are spectrally
classified as quasars\footnote{By spectral classification as a QSO, we mean a
  spectrum with broad emission lines, i.e., classified as a type~1
  (unobscured) AGN.  There is no consideration of luminosity or (narrow-)line
  ratios.\label{ftn:qso-definition}} and some as galaxies.

\begin{figure*}
\includegraphics[width=\doublecolsize\textwidth]{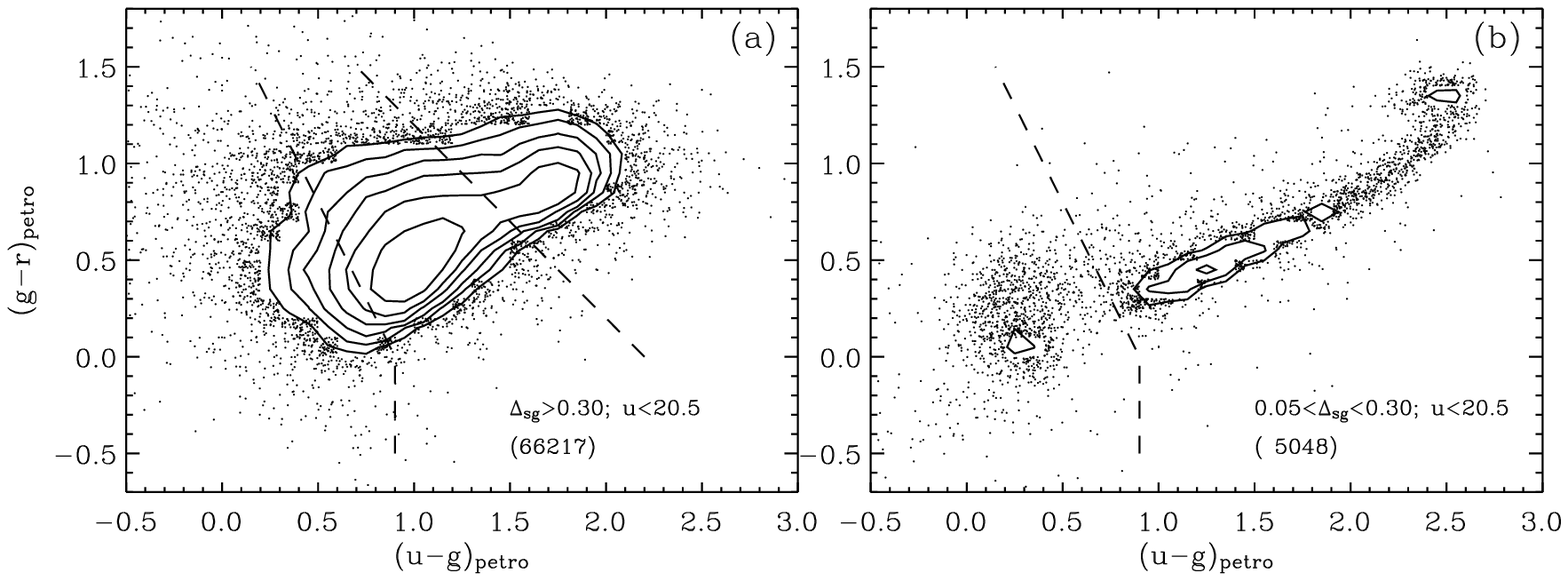}
\caption{Observed color-color plots 
  for (a) strongly resolved sources and (b) weakly resolved sources. The
  kinked dashed lines represent Eqs.~\ref{eqn:qso2} and~\ref{eqn:qso3}.
  Objects to the left of these could be selected by standard QSO selection 
  (10\% and 76\% of those observed are spectrally classified as QSO, 
  for each panel, respectively). 
  The straight dashed line [in Panel (a)] represents an
  approximate division of the bimodality in the galaxy distribution
  ($u-r=2.2$). Galaxies to the right are generally early types and those to
  the left are generally late types. The solid lines represent
  logarithmically-spaced density contours with 4 contours per factor of 10.
  Note that, for this plot, objects were restricted to those with ${\rm
    S/N}>5$ in $u$, $g$ and $r$, and with $u_{\rm extinction}<1$.}
\label{fig:color-color}
\vspace{0.6cm}
\includegraphics[width=\singlecolsize\textwidth]{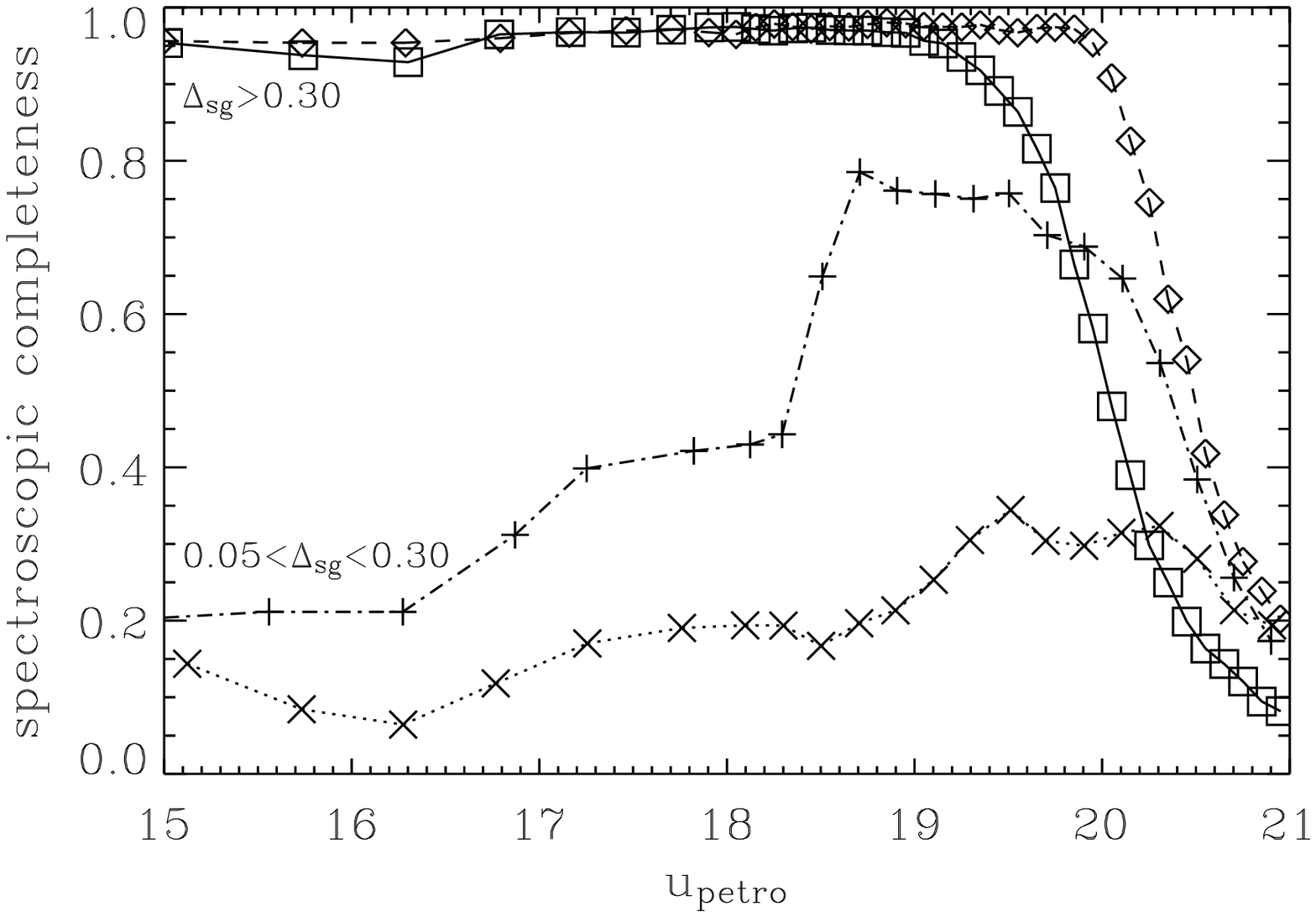}
\caption{Average completeness versus $u$-band magnitude 
  for different types of objects. The solid and dashed lines represent
  strongly resolved objects that are bluer and redder than $u-r=2.2$,
  respectively (Fig.~\ref{fig:color-color}a); the dash-and-dotted and dotted
  lines represent weakly resolved objects that are bluer and redder than the
  approximate QSO cut, respectively (Eqs.~\ref{eqn:qso2} and~\ref{eqn:qso3};
  Fig.~\ref{fig:color-color}b).}
\label{fig:completeness}
\vspace{0.6cm}
\includegraphics[width=\doublecolsize\textwidth]{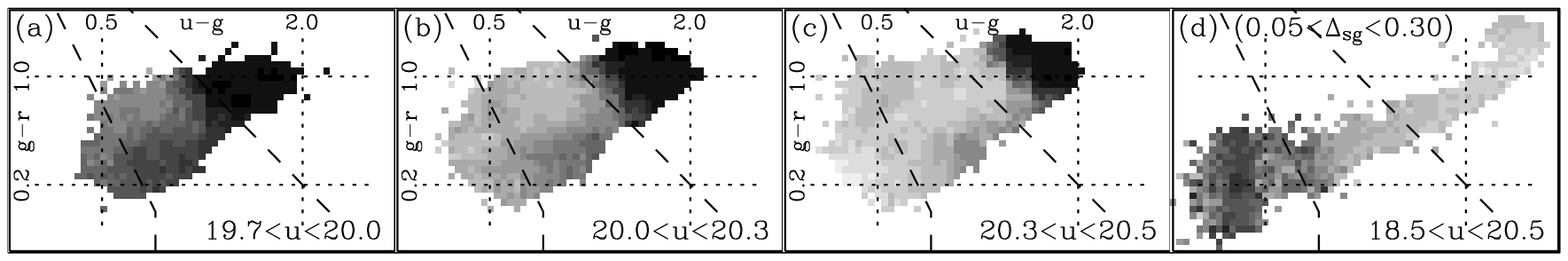}
\includegraphics[width=0.5\textwidth]{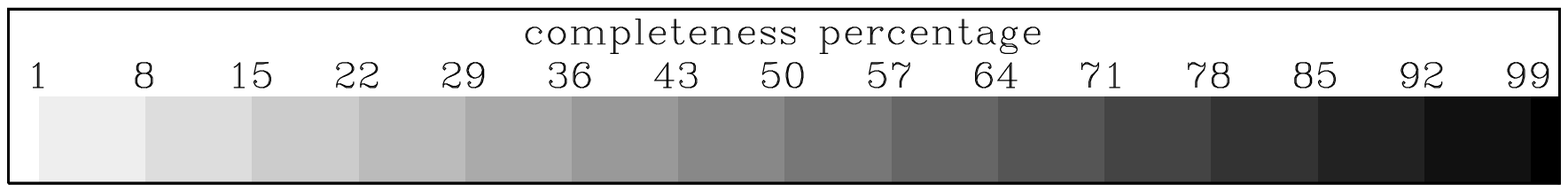}
\caption{Completeness as a function of $g-r$ versus $u-g$ for 
  (a--c) strongly resolved sources with various magnitude ranges and (d)
  weakly resolved sources.  Note there are no areas with $<1$\% completeness
  so white areas represent $<5$ objects per $0.05\times0.05$ color bin [$<2$
  objects in Panel (d)] . The dashed lines represent the same color cuts as in
  Fig.~\ref{fig:color-color}, which shows the number densities.}
\label{fig:comp-maps}
\end{figure*}

To determine completeness corrections, we divide the sample into bins using
four variables: $\sgsep$, $u$, $u-g$, and $g-r$.  These are related to the
primary selection variables described in \S~\ref{sec:spec-target}.  First,
galaxies are divided into two samples based on a cut in $\sgsep$ at 0.3 (near
the limit for MAIN selection).  Figure~\ref{fig:color-color} shows $u-g$
versus $g-r$ distributions for the strongly resolved and weakly resolved
sources (now divided at 0.3).  The kinked dashed lines show the limit for QSO
selection of extended sources \citep{richards02} while the straight dashed
line shows a cut at $u-r=2.2$, which approximately divides the bimodality in
the galaxy distribution \citep{strateva01}. The point to note is that the
completeness will vary with position in this diagram, because of cuts in $r$,
$g$ or associated colors. In addition, photometric redshift selection will
also depend strongly on these colors because the 400-nm break moves through
the $g$-band over the redshift range 0.0 to 0.3. 

After dividing the sample using $\sgsep$, for $u>19.7$, the strongly resolved
objects are divided into 0.1-mag bins and the weakly resolved into 0.2-mag
bins.  At brighter magnitudes, wider bins are used.  These magnitude bins are
further divided in $u-g$ and $g-r$ with from $2\times2$ to $16\times8$ color
bins such that there are a minimum of fifty objects per final bin. Note that
we have not included SB as a variable which would be necessary for studying
bivariate distributions involving SB (e.g.\ 
\citealt{cross01,blanton03broadband}).

Figure~\ref{fig:completeness} shows the average completeness as function of
magnitude for four groups divided by $\sgsep$ and by color. The completeness
is higher for the strongly resolved objects, with the redder objects having
higher completeness because of MAIN selection. In the weakly resolved group,
the bluer objects have higher completeness because of QSO selection.  For the
group dominated by late-type galaxies (solid line in
Fig.~\ref{fig:completeness}), the average completeness is 88\% at $u=19.5$
dropping smoothly to 18\% at $u=20.5$. Above this limit, the completeness
drops below 2\% for some bins ($\sgsep>0.3$).  Figure~\ref{fig:comp-maps}
shows completeness as a function of color.

For the SDSS $u$GS catalog, we use the following selection criteria from the
{\em coadded imaging catalog} (\S~\ref{sec:coadd-imaging}):
\begin{eqnarray}
\upetro & < & 20.5 \label{eqn:ugs1} \\
\gpetro & < & 21.0 \label{eqn:ugs2} \\
\sgsep  & > & 0.05 \label{eqn:ugs3} \\
u_{\rm extinction} & < & 1.4 \label{eqn:ugs4} \: .
\end{eqnarray}
The later cut on MW extinction only excludes objects over
$\approx0.7{\rm\,deg}^2$ around RA of 58.4\degr.\footnote{The area lost to
  bright stars is also small ($<1$\,\sqdeg\ in total), e.g., around the $V=3$
  star HR8414 and the globular cluster M2. We assume the area of the survey is
  275\,\sqdeg\ for calculating galaxy number and luminosity densities.} We
have not included an explicit SB limit because the production of the coadded
catalog (\S~\ref{sec:coadd-imaging}) is robust to low-SB artifacts. However,
there is an implicit limit because of the pipeline reduction
\citep{blanton04}. This does not affect our results significantly, which are
focused on the luminosity density (LD) measurements.

The catalog includes 74901 objects, of which, 45839 have been observed
spectroscopically with SDSS (see Table~\ref{tab:spec-select} for breakdown by
program), and 104 with the 2dFGRS; the average completeness is 61\%.  The
limit has been extended to 20.5 rather than stopping at the 19.8/20.0 limit of
the $u$-band selection (item~\ref{itm:u-select} of \S~\ref{sec:spec-target})
because pipeline flux changes and the other selections allow this. In other
words, all galaxy spectral types are sufficiently sampled spectroscopically as
faint as $u\approx20.5$.  Note that even at this limit, the redshift
reliability is very high ($\approx99$\%) because galaxies are either bright
enough in the visible ($\rpetro<19$) or they very likely have strong emission
lines. We assume that all the measured redshifts (D.~Schlegel et al.\ 
pipeline) are correct for calculating the LFs.

\section{Galaxy Luminosity functions}
\label{sec:lf}

From the $u$GS catalog (Eqs.~\ref{eqn:ugs1}--\ref{eqn:ugs4}), galaxies are
selected with spectroscopic redshifts in the range $0.005<z<0.3$ and with
magnitudes in the range $14.5<u<20.5$. In addition, we remove weakly resolved
sources ($\sgsep<0.3$) that are spectrally classified as a QSO (see
footnote~\ref{ftn:qso-definition}). In other words, only galaxies where the
integrated visible flux is dominated by stellar light are included. These
selections produce a sample of 43223 galaxies. Note that compact non-QSO
galaxies ($\sgsep<0.3$) contribute $<1$\% to the LD.

Following \citet{blanton03ld}, we $k$-correct to the rest-frame band
equivalent to the observed $u$-band at $z=0.1$, which is called the \usband\ 
band ($\lambda_{\rm eff}\approx322\,{\rm nm}$; ${\rm FWHM}\approx53\,{\rm
  nm}$). The absolute magnitude on the AB system \citep{OG83} is given by
\begin{equation}
M_{322} = \upetro - k_{0.1u,u} - 5 \log (D_L / 10\,{\rm pc}) - 0.04
\end{equation}
where: $D_L$ is the luminosity distance for a cosmology with \cosmos\ =
(0.3,0.7) and $H_0 = (h_{70}) {\rm\,70\,km\,s^{-1}\,Mpc^{-1}}$, and
$k_{0.1u,u}$ is the k-correction using the method of \citet{blanton03kcorr}
(see, e.g., \citealt{hogg02kcorr} for a general definition of the
$k$-correction). The $-0.04$ term is the estimated correction from the SDSS
$u$-band system to an AB system \citep{abazajian04}.\footnote{The SDSS
  photometry was originally designed to be calibrated to an AB scale but
  because of filter variations between the natural system of the telescope and
  the standard star system it has been modified \citep{abazajian03}.}

To calculate the galaxy LFs, we divide the sample into
0.02-redshift slices and 0.2-magnitude bins. For each bin, the LF is 
then given by
\begin{equation}
 \phi_M \, \dd M = \sum_i \frac{1}{{C_i}\,{V_{{\rm max},i}}}
\end{equation}
where: $C$ is the spectroscopic completeness (\S~\ref{sec:completeness}), and
$V_{\rm max}$ is the comoving volume over which the galaxy could be observed
\citep{Schmidt68} {\em within} the redshift slice and within the magnitude
limits of $14.5<u<20.5$.  Each redshift slice is cut in absolute magnitude so
that the sample is nearly volume limited ($0.9<V_{\rm max}/V_{\rm slice}\le1$
for $\sim95$\% of the galaxies).  The volumes of each slice range from
$2\times10^4{\rm\,Mpc}^3$ ($z=0.005$--0.02) to $8\times10^6{\rm\,Mpc}^3$
($z=0.28$--0.30).

\begin{figure*}
\includegraphics[width=\doublecolsize\textwidth]{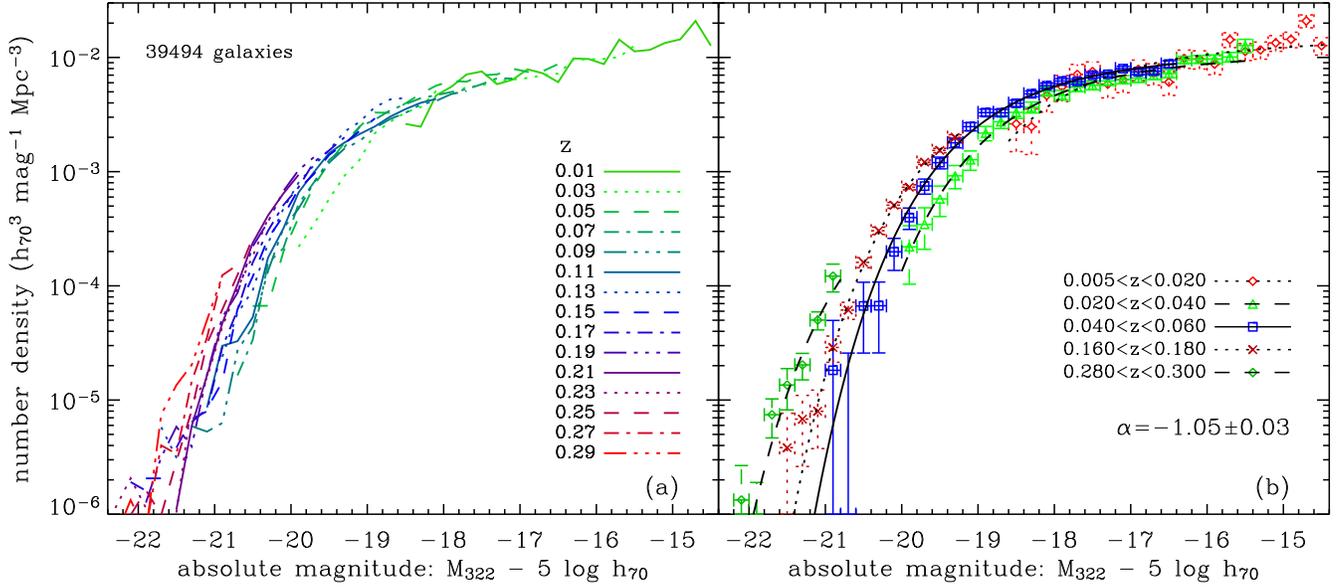}
\caption{Near-UV \usband\ LFs for galaxies with $0.005<z<0.3$.
  The lines represent different redshift slices. The binning is 0.02 in
  redshift and 0.2 in magnitude. There is a clear increase in the comoving
  density of luminous galaxies with increasing redshift.  Panel (a) shows the
  binned functions, while Panel (b) shows the binned-function error bars and
  Schechter fits for the three lowest redshift slices and two others (fixed
  $\alpha$).  The full sample over the redshift range includes 43223 galaxies,
  which is reduced to 39494 by making the slices nearly volume limited.}
\label{fig:lf}
\end{figure*}

Figure~\ref{fig:lf}(a) shows the binned galaxy LFs for the 15 redshift slices
out to $z=0.3$. The dominant effect is that the function shifts toward higher
luminosities with increasing redshift.

The general form of the LFs can be parameterized using the \citet{Schechter76}
function, which is given by
\begin{equation}
\phi_L \, \dd L = \phi^{*} \left( \frac{L}{L^{*}} \right)^\alpha 
e^{-L/L^{*}} \frac{ \dd L }{ L^{*} }
\end{equation}
where: $\phi_L \, \dd L$ is the comoving number density of galaxies with
luminosity between $L$ and $L + \dd L$; $L^{*}$ is the `characteristic
luminosity' (for the exponential cutoff); \phistar\ is the `characteristic
number density'; and $\alpha$ is the `faint-end slope'. This equation is
converted to magnitude form, with \mstar\ as the `characteristic magnitude',
and the slope is then $-0.4(\alpha+1)$ in a log number density versus
magnitude plot. We use this parametric form to quantify the evolution in terms
of \mstar\ and \phistar.

Before quantifying the evolution, we determine the faint-end slope.  The
density of galaxies fainter than $-17$ is only determined for the three lowest
redshift slices. On the assumption that the faint-end slope does not vary, we
fit the best-fit slope over these lowest slices ($z<0.06$). The Schechter fits
are shown in Fig.~\ref{fig:lf}(b).  The best-fit faint-end slope is given by
\begin{equation}
\alpha = -1.05 \pm 0.03 \, [\pm 0.07]
\end{equation}
where: the first uncertainty is the standard error from the Poisson noise;
and the second uncertainty is an estimate of the systematic error, obtained
by comparing with the best fits using combinations of three out of the four
lowest redshift slices (Fig.~\ref{fig:alphas}). These variations may reflect
the fact that the Schechter function is not a perfect match to the LFs (and
therefore depends on the magnitude range of the fitting), there may be
systematic uncertainties because of photometric deblending or astrophysical
changes because of large-scale structure.

Assuming that the faint-end slope does not vary significantly with redshift,
we can then characterize with more precision the evolution of the exponential
cutoff shown clearly in Fig.~\ref{fig:lf}.  To do this, we restrict the
fitting to $\alpha$ between $-1.15$ and $-0.95$.  For each redshift slice,
\mstar\ and \phistar\ are fitted marginalizing over the allowed range of
$\alpha$ (Table~\ref{tab:schechter-fits}). For the four highest redshift
slices ($z>0.22$), $\log\phi^{*}$ is constrained to be greater than $-2.2$;
and for the lowest redshift ($z<0.02$), \mstar\ is constrained to be brighter
than $-18.4$. This is because these parameters are not realistically fit in
these regimes.

\begin{figure*}
\includegraphics[width=\doublecolsize\textwidth]{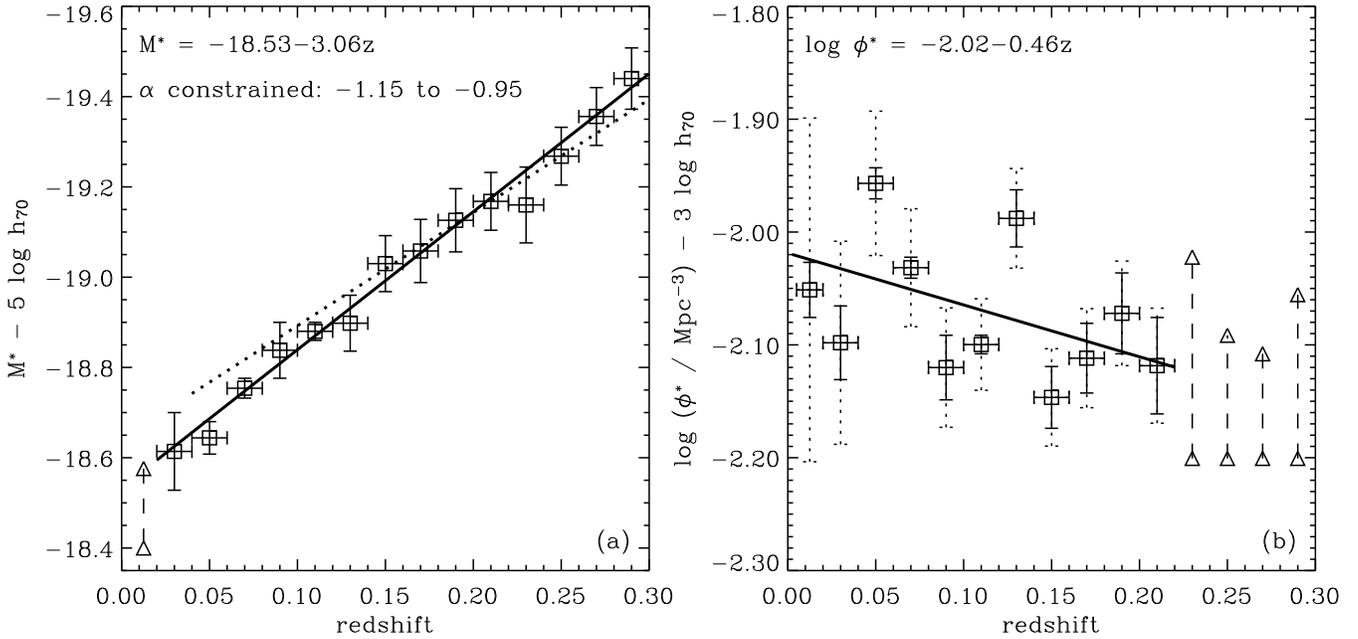}
\caption{Evolution in the \usband\ LFs 
  characterized by the Schechter parameters \mstar\ and \phistar.  The
  faint-end slope $\alpha$ is assumed to be in the range $-1.15$ to $-0.95$.
  The squares with solid error bars represent standard errors while the dotted
  error bars [in Panel (b)] represent standard errors plus an error added in
  quadrature to account for cosmic variance (Appendix~\ref{sec:lf-more}). The
  triangles with dashed lines represent the 1-$\sigma$ ranges when the fitting
  if restricted to $M^{*}<-18.4$ or $\log\phi^{*}>-2.2$.  The thick solid
  lines represent fits to the evolution over the range shown
  (Table~\ref{tab:evolve-fits}). The dotted line [in Panel (a)] represents a
  fit with fixed $\alpha=-1.05$.}
\label{fig:evolve-mstar}
\end{figure*}

Figure~\ref{fig:evolve-mstar} shows the evolution in the Schechter parameters.
There is a highly significant detection of evolution in \mstar\ with a slope
of $-3.1\pm0.2$ per redshift. This means that the exponential cutoff becomes
more luminous by about 0.9 mag between redshifts 0.0 and 0.3.  There is a
marginal detection of evolution in \phistar, which is more strongly affected
by cosmic variance.  This is only a 1.5-$\sigma$ detection using the redshift
slices up to 0.22. The fits to the evolution are given in
Table~\ref{tab:evolve-fits}.  Note that the results (for \mstar\ in
particular) depend significantly on assumptions about $\alpha$, for example,
for $\alpha=-1.05$ strictly fixed we obtain an \mstar\ evolution slope of
$-2.5\pm0.1$.  Thus, \mstar\ gets more luminous by $0.8\pm0.1$ mag ($z=0.0$ to
0.3) depending on the details of the LF evolution.

The lowest redshift slice was not included in the \mstar\ evolution fit.  This
was to avoid problems with measuring bright nearby galaxies using the SDSS
pipelines.  Large galaxies may be too strongly deblended (SB fluctuations are
more significant at low redshift), or the Petrosian aperture may not be large
enough.  In these cases, the flux of the galaxies will be underestimated.

\begin{table*}
\caption{Straight line fits to evolution in the Schechter parameters
  and the luminosity density}
\label{tab:evolve-fits}
\begin{center}
\begin{tabular}{lrrrrrr} \hline
parameter (322-nm band) & redshift & \multicolumn{3}{c}{line intercepts} 
& line slope & $\beta$ value\rlap{$^a$} \\
& fitting range & ($z=0.0$) & ($z=0.1$) & ($z=0.2$) & (per unit $z$) 
& \\
&&&&&& \\
$M^{*} - 5 \log h_{70}$ (mag) ..............\rlap{$^b$}
&0.02--0.30&$-18.53\pm0.02$&$-18.84\pm0.01$&$-19.15\pm0.02$&$ -3.06\pm0.18$&$ 3.3\pm0.2$\\
$\log(\phi^{*}\,/\,{\rm Mpc}^{-3}) - 3 \log h_{70}$ .....
&0.00--0.22&$ -2.02\pm0.04$&$ -2.06\pm0.02$&$ -2.11\pm0.03$&$ -0.46\pm0.31$&$-1.2\pm0.8$\\
$j + 2.5 \log h_{70}$ (mag ${\rm Mpc}^{-3}$) ...
&0.00--0.30&$-13.47\pm0.07$&$-13.68\pm0.03$&$-13.89\pm0.06$&$ -2.12\pm0.55$&$ 2.2\pm0.6$\\
&&&&&& \\
$M^{*} - 5 \log h_{70}$ (mag) ..............\rlap{$^c$}
&0.04--0.30&$-18.64\pm0.02$&$-18.89\pm0.01$&$-19.14\pm0.01$&$ -2.51\pm0.12$&$ 2.7\pm0.1$\\
$\log(\phi^{*}\,/\,{\rm Mpc}^{-3}) - 3 \log h_{70}$ .....
&0.00--0.22&$ -2.05\pm0.04$&$ -2.08\pm0.02$&$ -2.11\pm0.02$&$ -0.27\pm0.29$&$-0.7\pm0.8$\\
$j + 2.5 \log h_{70}$ (mag ${\rm Mpc}^{-3}$) ...
&0.00--0.30&$-13.53\pm0.06$&$-13.71\pm0.03$&$-13.89\pm0.03$&$ -1.80\pm0.37$&$ 1.9\pm0.4$\\
\hline
\end{tabular}
\end{center}
\begin{flushleft}
$^a$Fit to linear measurements ($L^{*}$, \phistar\ and $\rho_L$), with a
function $\propto(1+z)^\beta$.\newline 
$^b$For the first set of fits, the results were marginalized over the
faint-end slope: $\alpha$ in the range $-1.15$ to $-0.95$. By comparison with
fixed $\alpha$ results, estimates of the systematic uncertainties are 0.05 in
\mstar, 0.02 in $\log\phi^{*}$ and 0.04 in $j$ for the intercepts.\newline 
$^c$For the second set of fits, the faint-end slope was fixed: $\alpha=-1.05$.
Note here the 0.02--0.04 slice is an outlier in \mstar\ 
(Table~\ref{tab:schechter-fits}), which may reflect photometric errors or
large-scale structure variations.
\end{flushleft}
\end{table*}

\begin{figure}
\includegraphics[width=\singlecolsize\textwidth]{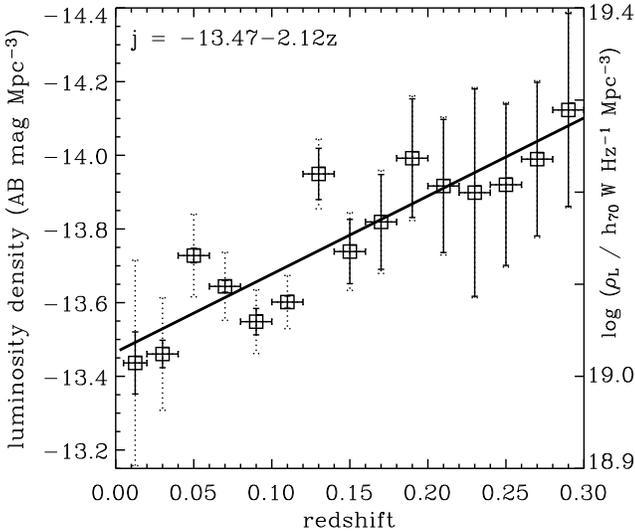}
\caption{Evolution in the \usband\ comoving LD.  
  See Fig.~\ref{fig:evolve-mstar} for symbol and line-style details.}
\label{fig:lum-dens}
\end{figure}

The Schechter function allows for an estimate of the total comoving LD
assuming that the function remains valid outside the magnitude range of the
fitting.  This LD is given in magnitudes per cubic megaparsec by
\begin{equation}
  j = M^{*} - 2.5 \log 
  \left[ (\phi^{*}/{\rm Mpc}^{-3}) \, \Gamma_{\rm f} (\alpha + 2) \right]
\label{eqn:lum-dens-schechter}
\end{equation}
where $\Gamma_{\rm f}$ is the gamma function; and in linear units by
\begin{equation}
  \rho_L = 10^{(34.1-j)/2.5} {\rm\:\:W\,Hz^{-1}\,Mpc^{-3}}
\end{equation}
from $j$ in AB mag ${\rm Mpc}^{-3}$.
If we assume that \phistar\ and $\alpha$ are
constant then the evolution in \mstar\ also represents the evolution in the
comoving LD.  However, we cannot rule out contributions from variations in
\phistar\ (or $\alpha$) to the evolution.  Figure~\ref{fig:lum-dens} shows the
evolution in the LD that includes the variation in \phistar.  Parameterizing
the evolution as
\begin{equation}
\rho_L \propto (1+z)^{\beta}
\label{eqn:l-evolve}
\end{equation}
as per for example \citet{lilly96}, then we obtain $\beta=2.2\pm0.6$ for the
322-nm LD. In \S~\ref{sec:evolve-ld}, we combine these measurements with
COMBO-17 at $0.2<z<1.2$ \citep{wolf03}.

\section{Discussion}
\label{sec:discuss}

\subsection{Comparison with $z\sim0.1$ luminosity functions}
\label{sec:compare-lowz}

The first test we make is to compare with the SDSS results of
\citet{blanton03ld} that used the same band.  For the \usband\ LF, their
sample included 22020 galaxies with $u<18.4$ in the redshift range
$0.02<z<0.14$.  The $u$ limit was chosen in order to avoid significant bias
from the $r<17.8$ MAIN selection, i.e., because 99\% of galaxies have
$u-r\ga0.6$ [Fig.~\ref{fig:color-color}(a)].  The fitting used a maximum
likelihood method, with a general shape for the LF, that incorporated
luminosity and number evolution.  The results obtained (by fitting a Schechter
function) were $M^{*} = -18.70\pm0.03$, $\alpha=-0.92\pm0.07$ and
$j=-13.71\pm0.14$ at $z=0.1$ (after converting to $H_0=70$). The faint-end
slope and LD are in 2-$\sigma$ statistical agreement with our results while
\mstar\ is not ($0.14$ mag difference).  However, if we set $\alpha$ to be
$-0.92$ in our analysis, the discrepancy is reduced. In addition, we used
fixed 0.2-magnitude bins whereas they used multiple-Gaussian fitting. Note
also that there are less than 4000 galaxies in common between the two data
sets ($<20$\% of theirs and $<10$\% of ours).

Parameters were used to include evolution in the luminosity and number density
\citep{lin99}. In terms of the evolution fitting given in
Table~\ref{tab:evolve-fits}, \citeauthor{blanton03ld}'s results can be
considered as $-4.2\pm0.9$ for \mstar\ and $+1.3\pm1.3$ for $\log \phi^{*}$
(both per unit $z$). These evolution parameters are in agreement with our
results within 2-$\sigma$.

\begin{figure}
\includegraphics[width=\singlecolsize\textwidth]{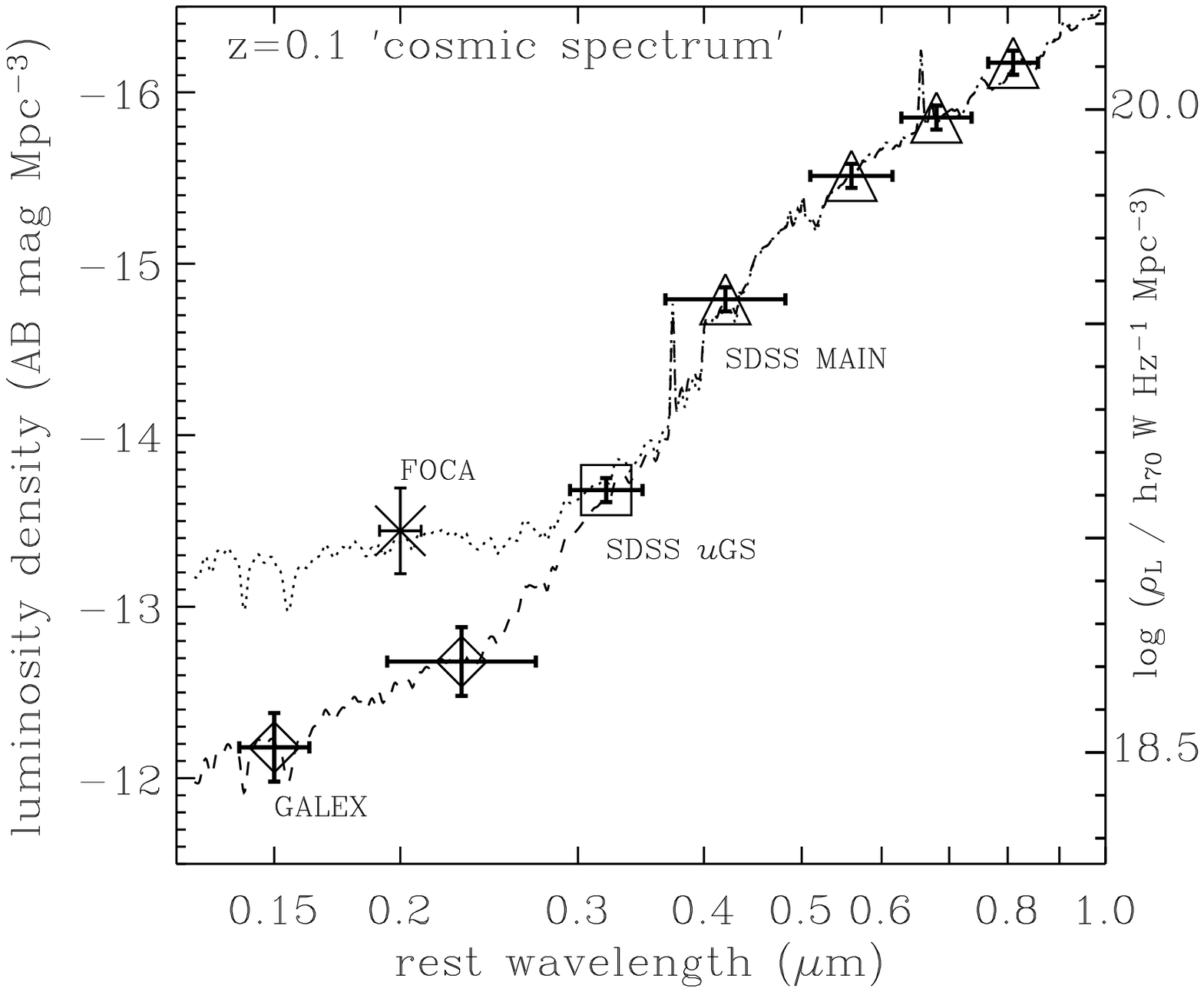}
\caption{Luminosity densities at $z=0.1$ from SDSS, FOCA and GALEX.
  The square represents the \usband\ result of this paper; the triangles, the
  $^{0.1}griz$ results of \citet{blanton03ld}; the cross, the FOCA result of
  \citet{sullivan00}; and the diamonds, the GALEX results of
  \citet{budavari05}.  The horizontal bars represent the FWHM of each band.
  The best-fit \citet{FR97,FR99} population-synthesis model from the fitting
  of \citet{BG03} is shown by the dotted line; and a new best-fit model that
  uses the GALEX results is shown by the dashed line.}
\label{fig:example-fit}
\end{figure}

Using the $u$GS, we have improved on the accuracy of the \usband\ LF and
evolution compared to using the main galaxy sample selected in the $r$ band
(which was of course a strong motivation for this survey).  For example, the
\usband\ LD is now known to comparable accuracy as the
$^{0.1}griz$ bands measured by \citet{blanton03ld}.
Figure~\ref{fig:example-fit} shows the $z=0.1$ luminosity densities from the
SDSS. The uncertainties are of order 5--10\% because of absolute calibration,
SB selection effects, conversions to total galaxy magnitudes and estimates of
the effective survey areas.

Figure~\ref{fig:example-fit} also shows the luminosity densities from the
ballon-borne FOCA telescope \citep{sullivan00} and from the GALEX space
telescope \citep{budavari05}. The former was derived from 2.2\,\sqdeg\ over
$0.0<z<0.4$ and the latter from 44\,\sqdeg\ over $0.07<z<0.13$.  The earlier
result is significantly higher in LD.  Also, \citeauthor{sullivan00} obtained
a faint-end slope of $-1.5\pm0.1$, which is not in agreement with our
measurements, while \citeauthor{budavari05} obtained $-1.1\pm0.1$, which is in
good agreement. In Fig.~\ref{fig:example-fit}, we also show two spectral
models for the UV to near-IR `cosmic spectrum' from the fitting of
\citet{BG03} (using PEGASE models; \citealt{FR97,FR99}).  We use the new fit
to the GALEX plus SDSS measurements to obtain a correction from the \usband\ 
band to a $280\pm20$\,nm band (Fig.~\ref{fig:filters}), which we use to
compare with higher redshift measurements of rest-frame UV densities.  This is
given by
\begin{equation}
  j_{280} \approx j_{322} + 0.45 \: .
\label{eqn:j-conv}
\end{equation}
A similar correction is determined if we use the $k$-correction templates from
\citet{blanton03kcorr}, i.e., $k$-correcting to $\sim$ $^{0.3}u$ band, but
this is less reliable because the $z\sim0.1$ LF then requires significant
wavelength extrapolation from the observed bands.

Our low-redshift \uband\ results (Table~\ref{tab:schechter-fits}) are also in
good agreement with that determined in nearby clusters by \citet*{CMZ04}. They
found $M_{U}^{*} = -18.9\pm0.3$ (converting to AB mag) and
$\alpha_{U}=-1.1\pm0.2$.

\subsection{Evolution in comoving luminosity densities}
\label{sec:evolve-ld}

\begin{figure*}
\includegraphics[width=\doublecolsize\textwidth]{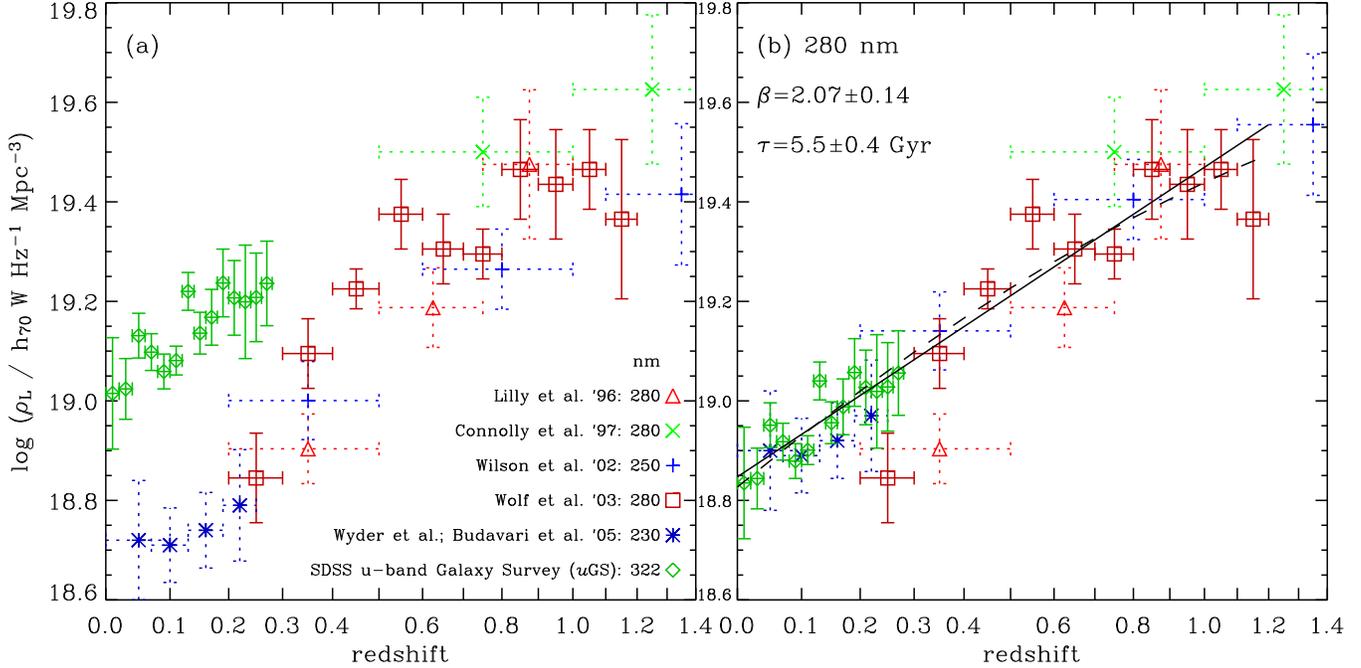}
\caption{Evolution in the near-UV comoving luminosity densities.
  Various data measurements are shown by the symbols. Panel (a) shows the data
  at wavelengths 230, 250, 280 and 322\,nm; while Panel (b) shows the data
  adjusted to 280\,nm where necessary.  The corrections in $\log\rho_L$ were
  assumed to be $-0.18$, $+0.18$ and $+0.14$ for the \usband, GALEX NUV and
  \citeauthor{wilson02}\ results, respectively (from the dashed-line fit in
  Fig.~\ref{fig:example-fit}).  Thus, the agreement between the results from
  this paper and from GALEX in Panel (b) is by construction.  The best fit to
  the data from this paper and from COMBO-17 (fig.~19 of \citeauthor{wolf03})
  is shown by a solid line using the parameterization of
  Eq.~\ref{eqn:l-evolve}, and by a dashed line using an exponentially
  increasing LD with look-back time. The $x$-axis is linear in $\log(1+z)$.}
\label{fig:ld-z}
\end{figure*}

In order to compare with redshifts out to $z\sim1$, we take near-UV LD
measurements from the literature
\citep{lilly96,connolly97,wilson02,wolf03,budavari05,wyder05} and convert them
to the \cosmos\ = (0.3,0.7) cosmology. This was done by making a comoving
volume correction over the redshift range of each measurement and a luminosity
correction for the midpoint redshift. Figure~\ref{fig:ld-z} shows these
measurements and our results converted to 280\,nm, which show an increase in
the LD with redshift. Our results and COMBO-17's results (both shown using
solid-line error bars) cover a significantly larger volume than the other
results (excepting the GALEX results). We fit to these SDSS and COMBO-17
results, and obtain
\begin{equation}
\beta_{280} = 2.07 \pm 0.14 \, [\pm 0.10]
\label{eqn:beta-280}
\end{equation}
where the first uncertainty is the standard error and the second is per 0.05
mag uncertainty in the Eq.~\ref{eqn:j-conv} conversion. This represents the
most accurate measurement of near-UV LD evolution to date and it rules out the
steep evolution found by \citet{lilly96} and is closer to the gradual rates
determined by \citet*{CSB99} and \citet{wilson02}. The Eq.~\ref{eqn:l-evolve}
fit is shown by a solid line in Fig.~\ref{fig:ld-z}(b), while a fit using
$\rho_L\propto\exp({t/\tau})$, where $t$ is the look-back time for our assumed
cosmology, is shown by a dashed line. The exponential timescale is given by
$\tau_{280}=5.5\pm0.4\,[\pm 0.3]{\rm\,Gyr}$.

In the absence of dust and chemical evolution, the 280-nm density evolution
corresponds closely to the SFR density. From a solar-metallicity population
synthesis model (with a $\beta\sim2$ star-formation history): $\beta_{\rm
  SFR}\sim\beta_{280}+0.1$ and $\tau_{\rm SFR} \sim\tau_{280}-0.2{\rm\,Gyr}$.
In other words, the evolution in the SFR is slightly steeper than that of the
near-UV LD because of relatively small contributions from evolved stellar
populations. Our results are consistent with that found using the far-UV by
\citet{schiminovich05}: $\beta_{150}=2.5\pm0.7$.

Many other SFR indicators have been used to trace the cosmic star formation
history, some of which are less sensitive to dust or can be corrected for
dust. From a compilation of UV, [O{\small II}], H$\alpha$, H$\beta$, mid-IR,
sub-mm, radio and X-ray measurements, \citet{Hopkins04} found that $\beta_{\rm
  SFR}=3.3\pm0.3$ for $z<1$ data when including SFR-dependent dust attenuation
corrections where necessary.  This is inconsistent with our result $\beta_{\rm
  SFR}=2.2\pm0.2$ that assumes no dust evolution.  The results can be
reconciled if the effective average attenuation at 280\,nm increases by
$0.8\pm0.3$ mag between $z=0$ and 1. This is a plausible increase because the
characteristic luminosities of rapidly star-forming galaxies are increasing
with redshift \citep{cowie96}; and more luminous galaxies have higher levels
of dust attenuation \citep{hopkins01}. However, we caution that our measured
dust increase is not independent of that idea because of the SFR-dependent
corrections used by \citeauthor{Hopkins04}. Bolometric OB-star luminosity
densities derived from far/near-UV and mid/far-IR wavelengths (with
corrections for AGN and evolved stellar populations) could be used for an
unambiguous measure of $\beta_{SFR}$.

\subsection{Evolution in luminosity functions}
\label{sec:evolve-lf}

While it is robust to compare our LD results with those of COMBO-17 using a
model cosmic-spectrum correction, it is not so straightforward for assessing
LF evolution.  Any color-magnitude relations and dispersion will affect the
number density and shape of the LFs.  Instead, the COMBO-17 data could be
analyzed to measure the LFs in the \usband\ or the more standard \uband\ band
(Appendix~\ref{sec:lf-more}). Other large surveys for which our results could
be compared are the VIMOS-VLT Deep Survey (VVDS; \citealt{lefevre04}) or the
DEEP2 survey \citep{davis03}.

\begin{figure}
\includegraphics[width=\singlecolsize\textwidth]{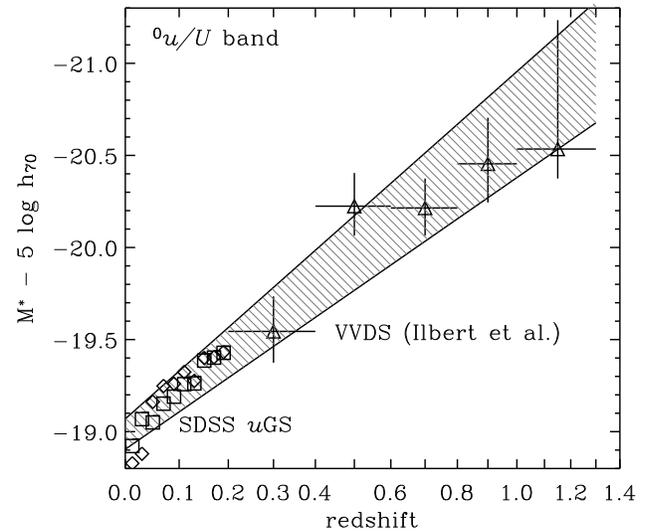}
\caption{Evolution in \mstar\ for the \uband\ LFs. 
  The squares and diamonds represent the results from this paper, marginalized
  over $\alpha=-1.15\pm0.10$ and fixed $\alpha=-1.15$, respectively (with
  errors approximately the size of symbols).  The triangles represent the
  results from \citet{ilbert04} with $\alpha$ allowed to vary except for the
  highest redshift slice. The sloped-line region shows the range between the
  lower and upper straight line fits.}
\label{fig:mstar-z}
\end{figure}

Comparing our \uband\ results with the VVDS results of \citet{ilbert04}, we
find \mstar\ brightens by 1.5 to 2 mag between $z=0$ and 1.
Figure~\ref{fig:mstar-z} shows the evolution in \mstar\ versus redshift. To
obtain upper and lower limits, we fit to the SDSS data ($\pm2\sigma$) and the
VVDS data ($\pm1\sigma$) allowing for the errors to all be in the same
direction. There is a suggestion that \mstar\ brightens more rapidly below
$z=0.5$ than above but the data are also consistent with a constant slope.

\section{Conclusions}
\label{sec:conclude}

We have analyzed a $u$-band selected galaxy survey, using spectroscopic
completeness corrections as a function of color and magnitude to account for
inhomogeneous selection (Figs.~\ref{fig:z-histos}, \ref{fig:completeness}
and~\ref{fig:comp-maps}). The main results are:
\begin{enumerate}
\renewcommand{\theenumi}{(\arabic{enumi})}
\item Testing star-galaxy separation (Fig.~\ref{fig:sgsep-z}), we find that
  compact galaxies that could be missed by MAIN selection contribute
  insignificantly to the stellar UV LD ($<1$\%).
\item The faint-end slope of the low-redshift \usband\ LF is near flat:
  $\alpha=-1.05\pm0.08$. This was obtained from the best fit over the three
  lowest redshift slices ($z<0.06$) with magnitudes from $-21.4$ to $-14.4$
  [Fig.~\ref{fig:lf}(b)].
\item The evolution in the LFs is dominated by a luminosity shift
  (Figs.~\ref{fig:lf} and~\ref{fig:evolve-mstar}), which can be characterized
  by a shift in \mstar\ of $-0.8\pm0.1$ mag between $z=0$ and 0.3.
\item In order to compare the UV LD evolution with COMBO-17 at higher
  redshifts \citep{wolf03}, we fit to LD measurements at $z=0.1$
  (Fig.~\ref{fig:example-fit}) including recent GALEX results of
  \citet{budavari05}, and obtain a correction to 280\,nm of $+0.45$ mag. Our
  LD measurements versus redshift then lineup remarkably well with those of
  COMBO-17 (Fig.~\ref{fig:ld-z}) and we find that the evolution can be
  parameterized by Eq.~\ref{eqn:l-evolve} with $\beta_{280}=2.1\pm0.2$. This
  is a significantly shallower evolution than that found using other SFR
  indicators and is consistent with an increase in average dust attenuation of
  $0.8\pm0.3$ mag between $z=0$ and 1.
\end{enumerate}
Future work with the $u$GS could include: assessing the UV LF evolution for
different classes of objects (e.g., AGN, late/early-type galaxies), measuring
the H$\alpha$ LF (modulo aperture and dust corrections) and studying
multi-variate distributions in a star-forming sample. These will enable a
quantification of contributions to the near-UV decline at $z\la0.2$ from
different galaxy populations and from dust and star formation.

\section*{Acknowledgments}

I.\,K.\,B.\ and K.\,G.\ acknowledge generous funding from the David
and Lucille Packard foundation.  T.\,B.\ acknowledges partial support
from the Hungarian Scientific Research Fund (OTKA) via grants T037548
and T047244. We thank Ryan Scranton for discussion concerning coadded
catalogs, Laurence Tresse for a preview of VVDS results, Andrew West
for discussion concerning sky-subtraction errors, Christian Wolf for
information on COMBO-17, and the anonymous referee for improvements to
the manuscript.

Funding for the creation and distribution of the SDSS Archive has been
provided by the Alfred P. Sloan Foundation, the Participating
Institutions, the National Aeronautics and Space Administration, the
National Science Foundation, the U.S. Department of Energy, the
Japanese Monbukagakusho, and the Max Planck Society. The SDSS Web site
is {\tt http://www.sdss.org/}.

The SDSS is managed by the Astrophysical Research Consortium (ARC) for
the Participating Institutions. The Participating Institutions are The
University of Chicago, Fermilab, the Institute for Advanced Study, the
Japan Participation Group, The Johns Hopkins University, the Korean
Scientist Group, Los Alamos National Laboratory, the
Max-Planck-Institute for Astronomy (MPIA), the Max-Planck-Institute
for Astrophysics (MPA), New Mexico State University, University of
Pittsburgh, Princeton University, the United States Naval Observatory,
and the University of Washington.

\appendix

\section{More details on the luminosity functions}
\label{sec:lf-more}

\begin{figure}
\includegraphics[width=\singlecolsize\textwidth]{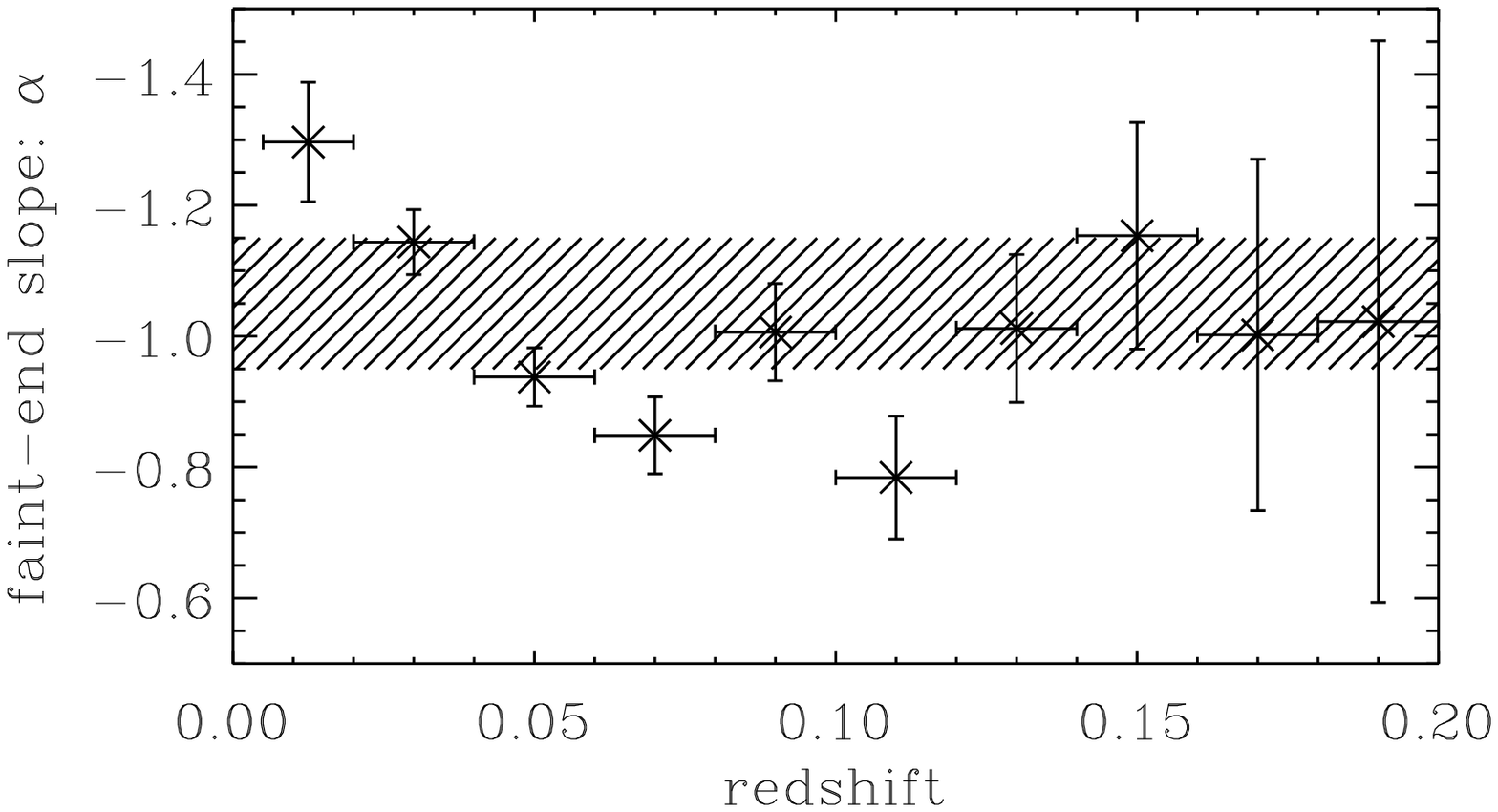}
\caption{Best-fit faint-end slopes for the \usband\ LFs. 
  The horizontal bars represent the redshift ranges while the vertical error
  bars represent the 1-$\sigma$ errors. The sloped-line region shows the
  allowed $\alpha$ range when fitting the other Schechter parameters. Note
  that we do not interpret the change in $\alpha$ from $z=0.0$ to 0.1 as being
  caused by galaxy evolution.  The change could be caused by the varying
  magnitude limits in the fitting (Table~\ref{tab:schechter-fits}). For the
  lowest redshift slices there is a higher weight from luminosities fainter
  than $\sim M^{*}+1$; whereas at $z\sim0.05$--0.1 the fitting is giving
  higher weight to the `knee' of the LF. In other words, the Schechter
  function is not a perfect match to the LF as there appears to be a change in
  slope around $M_{322}\sim-16.5$ (Fig.~\ref{fig:lf}). For the purposes of
  this paper, assessing the evolution in \mstar\ and $j$, we use the
  three-parameter Schechter function and compromise on $\alpha$.  Note also
  there could be systematic errors at low redshift because of deblending
  issues that increase the number of faint galaxies at the expense of bright
  galaxies.}
\label{fig:alphas}
\end{figure}

We use a straightforward \Vmax\ in slices approach for determining the LFs
(\S~\ref{sec:lf}). Schechter functions are fitted to the LFs using standard
least-squares routines.  The errors on each 0.2-mag bin are taken to be a
modified Poisson error where the variance without weighting would be $N+2$.
This is an appropriate variance for low number counts (when the expected value
is not known).  Figure~\ref{fig:alphas} shows the best-fit faint-end slopes
versus redshift. As there are significant degeneracies between \mstar\ and
$\alpha$, we considered a limited range in $\alpha$, $-1.15$ to $-0.95$, in
order to assess the evolution in \mstar\ and \phistar.
Table~\ref{tab:schechter-fits} gives the Schechter parameters for the redshift
slices along with some details of the LF fitting.

\begin{table*}
\caption{Luminosity function fitting and Schechter parameters $^a$}
\label{tab:schechter-fits}
\begin{center}
\begin{tabular}{|c|cc|ccc|cccc|} \hline
range of & \multicolumn{2}{c|}{$M_{322}$ limits} & no.~of 
& mean~of & mean~of & 
\multicolumn{4}{c|}{$M_{322}$ paras.\ marginalized over $\alpha=-1.05\pm0.10$}
\\
$z$ & bright & faint & galaxies & $1/C$ & $V_{\rm slice}/V_{\rm max}$ &
$-M^{*}$ & $-\log\phi^{*}$ & $-j$ & $1/f_{j,{\rm obs}}$ 
\\
&&&&&&&&&\\
0.005--0.02&$-18.6$&$-14.4$& 601&1.07&1.01&$<18.58$ $^b$ &$2.05\pm0.15$&$13.44\pm0.28$&1.32--1.55\\
 0.02--0.04&$-20.0$&$-15.4$&2421&1.17&1.01&$18.61\pm0.09$&$2.10\pm0.09$&$13.46\pm0.15$&1.06--1.14\\
 0.04--0.06&$-21.4$&$-16.4$&4738&1.15&1.01&$18.64\pm0.04$&$1.96\pm0.06$&$13.73\pm0.11$&1.09--1.13\\
 0.06--0.08&$-21.4$&$-17.0$&5334&1.23&1.01&$18.75\pm0.02$&$2.03\pm0.05$&$13.64\pm0.09$&1.17--1.21\\
 0.08--0.10&$-21.6$&$-17.6$&5058&1.25&1.01&$18.84\pm0.06$&$2.12\pm0.05$&$13.55\pm0.09$&1.33--1.42\\
 0.10--0.12&$-21.6$&$-18.0$&5143&1.34&1.03&$18.88\pm0.02$&$2.10\pm0.04$&$13.60\pm0.07$&1.46--1.51\\
 0.12--0.14&$-21.8$&$-18.4$&5549&1.47&1.04&$18.90\pm0.06$&$1.99\pm0.04$&$13.95\pm0.09$&1.81--2.05\\
 0.14--0.16&$-22.0$&$-18.8$&3549&1.51&1.04&$19.03\pm0.06$&$2.15\pm0.04$&$13.74\pm0.11$&2.25--2.64\\
 0.16--0.18&$-22.2$&$-19.2$&2719&1.49&1.06&$19.06\pm0.07$&$2.11\pm0.04$&$13.82\pm0.14$&2.95--3.74\\
 0.18--0.20&$-22.2$&$-19.6$&1987&1.48&1.03&$19.13\pm0.07$&$2.07\pm0.05$&$13.99\pm0.17$&4.35--5.85\\
 0.20--0.22&$-22.2$&$-19.8$&1259&1.74&1.06&$19.17\pm0.06$&$2.12\pm0.05$&$13.92\pm0.19$&5.46--7.62\\
 0.22--0.24&$-22.4$&$-20.2$& 467&1.71&1.04&$19.16\pm0.08$&$>2.02$ $^c$ &$13.90\pm0.29$&11.7--19.7\\
 0.24--0.26&$-22.4$&$-20.4$& 313&1.92&1.07&$19.27\pm0.06$&$>2.09$ $^c$ &$13.92\pm0.22$&16.0--24.0\\
 0.26--0.28&$-22.4$&$-20.6$& 220&2.17&1.04&$19.36\pm0.06$&$>2.10$ $^c$ &$13.99\pm0.21$&21.4--31.4\\
 0.28--0.30&$-22.4$&$-20.8$& 136&2.17&1.18&$19.44\pm0.07$&$>2.05$ $^c$ &$14.12\pm0.27$&29.6--47.9\\
\hline
\end{tabular}
\end{center}
\begin{center}
\begin{tabular}{|c|cccc|cccc|} \hline
range of & \multicolumn{4}{c|}{$M_{322}$ paras.\ with fixed $\alpha=-1.05$}
& \multicolumn{4}{c|}{$M_{355}$ paras.\ with fixed $\alpha=-1.05$\rlap{ $^d$}}
\\
$z$ & $-M^{*}$ & $-\log\phi^{*}$ & $-j$ & $1/f_{j,{\rm obs}}$
    & $-M^{*}$ & $-\log\phi^{*}$ & $-j$ & $1/f_{j,{\rm obs}}$
\\
&&&&&&&&\\
0.005--0.02&$<18.45$ $^b$ &$1.96\pm0.15$&$13.53\pm0.27$&1.50--1.64&$<18.86$ $^b$ &$2.00\pm0.15$&$13.85\pm0.27$&1.44--1.58\\
 0.02--0.04&$18.47\pm0.05$&$2.02\pm0.08$&$13.44\pm0.15$&1.04--1.11&$18.88\pm0.05$&$2.06\pm0.08$&$13.76\pm0.15$&1.04--1.11\\
 0.04--0.06&$18.76\pm0.03$&$2.02\pm0.06$&$13.75\pm0.11$&1.11--1.15&$19.16\pm0.03$&$2.04\pm0.06$&$14.08\pm0.11$&1.08--1.11\\
 0.06--0.08&$18.85\pm0.02$&$2.08\pm0.05$&$13.67\pm0.09$&1.20--1.24&$19.25\pm0.02$&$2.12\pm0.05$&$13.99\pm0.09$&1.20--1.24\\
 0.08--0.10&$18.87\pm0.02$&$2.14\pm0.05$&$13.57\pm0.08$&1.38--1.42&$19.26\pm0.02$&$2.16\pm0.05$&$13.88\pm0.08$&1.39--1.43\\
 0.10--0.12&$18.95\pm0.02$&$2.13\pm0.04$&$13.65\pm0.07$&1.54--1.58&$19.32\pm0.02$&$2.15\pm0.04$&$13.98\pm0.07$&1.56--1.60\\
 0.12--0.14&$18.91\pm0.02$&$2.00\pm0.04$&$13.96\pm0.07$&1.91--1.98&$19.28\pm0.02$&$2.01\pm0.04$&$14.28\pm0.07$&1.96--2.02\\
 0.14--0.16&$19.01\pm0.02$&$2.13\pm0.04$&$13.71\pm0.06$&2.32--2.43&$19.40\pm0.02$&$2.17\pm0.04$&$14.01\pm0.06$&2.36--2.48\\
 0.16--0.18&$19.06\pm0.03$&$2.11\pm0.04$&$13.81\pm0.07$&3.19--3.41&$19.40\pm0.03$&$2.10\pm0.04$&$14.18\pm0.07$&4.30--4.70\\
 0.18--0.20&$19.13\pm0.03$&$2.07\pm0.04$&$13.98\pm0.08$&4.76--5.27&$19.43\pm0.04$&$2.02\pm0.05$&$14.42\pm0.09$&7.57--8.73\\
\hline
\end{tabular}
\end{center}
\begin{flushleft}
$^a$See text of Appendix~\ref{sec:lf-more} for details. The units of \mstar, 
\phistar\ and $j$ are AB mag, ${\rm Mpc}^{-3}$ and AB mag ${\rm Mpc}^{-3}$,
respectively.\newline 
$^b$The \mstar\ limits shown for the lowest redshift slice are 1-$\sigma$
limits when \mstar\ is constrained to be brighter than $-18.4$ and $-18.8$ for
322\,nm and 355\,nm, respectively.\newline 
$^c$The \phistar\ limits for the redshift slices above 0.22 are 1-$\sigma$
limits when $\log\phi^{*}$ is constrained to be greater than $-2.2$.\newline 
$^d$The faint $M_{355}$ limits for the fitting were $(-0.2, -0.4, -0.6)$, with
respect to the $M_{322}$ limits, for the redshift ranges (0--0.06, 0.06--0.16,
0.16--0.20), respectively. The bright $M_{355}$ limits were $-0.2$ or $-0.4$
with respect to the $M_{322}$ limits. Beyond $z=0.2$, the \uband\ LFs are
significantly less reliable because of \Vmax\ and $k$-corrections, and a
$g$-based sample would be more appropriate.
\end{flushleft}
\end{table*}

The top half of the table gives the absolute magnitude ranges used in the
fitting, the number of galaxies, the mean of the correction factors and the
Schechter parameters for $\alpha$ marginalized over the range $-1.15$ to
$-0.95$. The column $1/f_{j,{\rm obs}}$ gives the 1-$\sigma$ range (not
including cosmic variance) of the inverse of the fraction of LD that is within
the magnitude limits, i.e., it represents the extrapolation factor from the
observed to total LD.  The \phistar\ and $j$ errors include an estimate of
cosmic variance, that is proportional to $V^{-0.3}$, added in quadrature to
the Poisson errors. This approximate power and the normalization were
estimated from fig.~3 of \citet{somerville04cv}. The cosmic variance on $j$ is
assumed to be less than that on \phistar\ because the specific SFR per galaxy
increases in low-density environments, which partly offsets changes in number
density.

The lower half of Table~\ref{tab:schechter-fits} gives the Schechter
parameters for fixed $\alpha=-1.05$ both for the \usband\ and the \uband\ 
band. From the \mstar\ and $j$ differences between the different bands, we
obtain
\begin{equation}
M_{355} \approx M_{322} - 0.35
\end{equation}
for the average galaxy, which is in good agreement with that determined from
the cosmic spectrum fit (Fig.~\ref{fig:example-fit}).

We provide these Schechter parameters to make comparisons with other surveys
but they are strictly only valid from about $M^{*}-2.5$ (or the bright limit)
to the faint limit of the fitting. At magnitudes brighter than about
$M^{*}-2.5$, there is a significant excess of galaxies above the Schechter fit
(from a composite LF with each redshift slice shifted by \mstar).

\section{Sky-subtraction corrections for SDSS {$u$}-band Petrosian fluxes}
\label{sec:sky-sub-corr}

\begin{figure}
\includegraphics[width=\singlecolsize\textwidth]{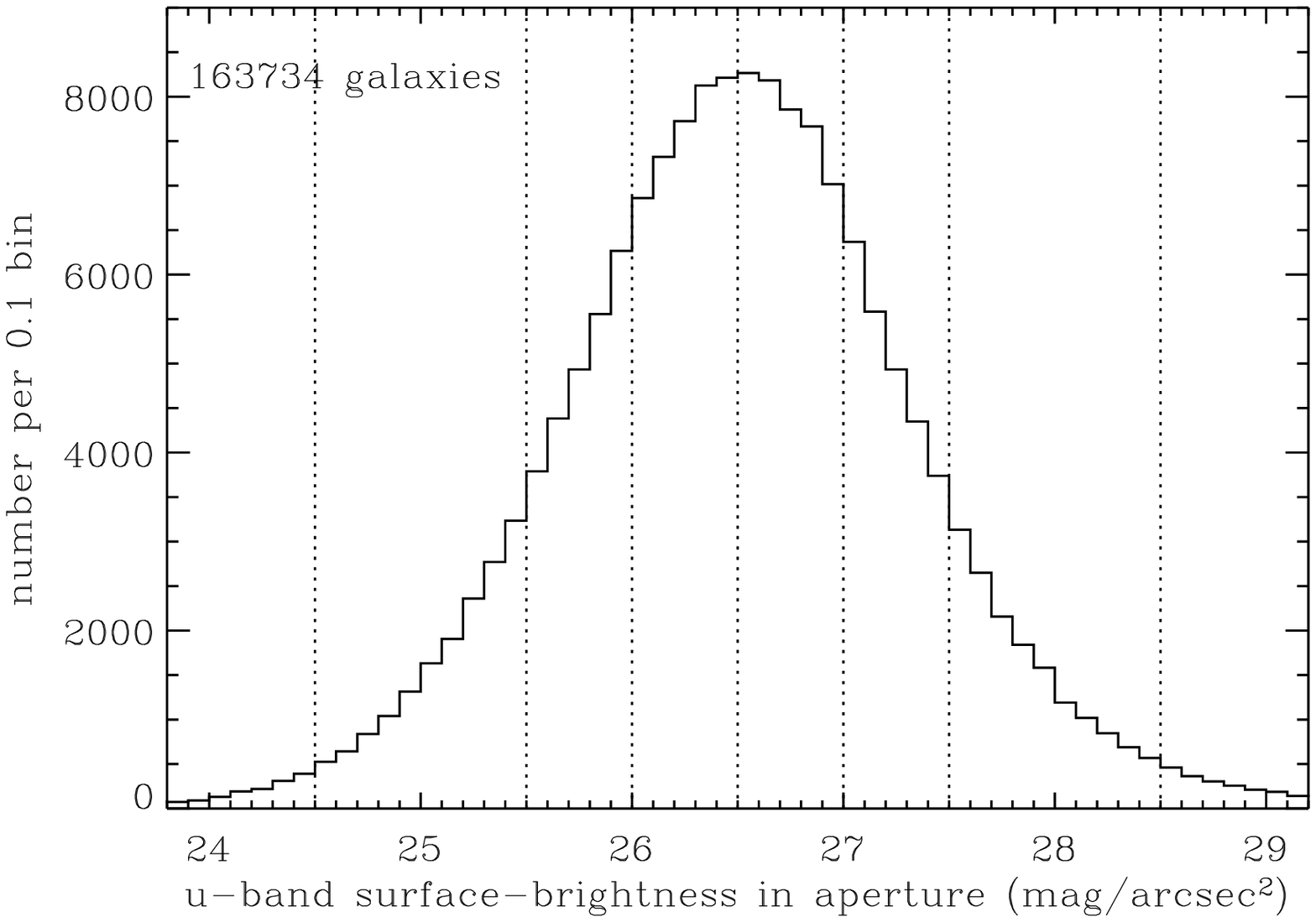}
\caption{Distribution of $u$-band surface brightnesses 
  for a galaxy sample with $\rpetro<19.4$. The vertical dotted lines show the
  positions of the cuts used to divide the sample for testing.  Note that the
  surface brightnesses are for the SDSS Petrosian aperture, which is two times
  the Petrosian radius in the $r$-band, and the values have not been corrected
  for MW extinction (unlike all other flux measurements in this paper).}
\label{fig:u-sp-aperture}
\end{figure}

The equivalent depths across the SDSS magnitudes for galaxy target selection
correspond to ($u,g,r,i,z$) $\approx$ (19.8, 18.6, 17.8, 17.4, 17.1).  In
other words, there are an equivalent number of galaxies per square degree
($\sim100$\,\persqdeg) for galaxies brighter than these limits in each of the
bands.  These also approximately correspond to the average spectral energy
distribution of galaxies near those magnitudes.  Comparing these limits (plus
average MW extinction) to the point-source 95\% completeness limits given in
\citet{stoughton02}, we obtain differences of (1.9, 3.4, 4.2, 3.8, 3.3).
Thus, the $u$-band galaxy measurements are, on average, the closest to the
survey imaging detection limit.  In addition, systematic errors can dominate
because scattered light can be significant in comparison with the sky flux
(which is less of a problem in the $z$-band, for example). This can affect
flat fielding and the zero point of the sky-subtracted frame and therefore is
most significant for galaxies with low surface brightnesses.  The distribution
of $u$-band surface brightnesses in the Petrosian aperture ($2\times$ the
Petrosian radius in the $r$-band) is shown in Fig.~\ref{fig:u-sp-aperture}
for a galaxy sample.

\begin{figure*}
\includegraphics[width=0.9\textwidth]{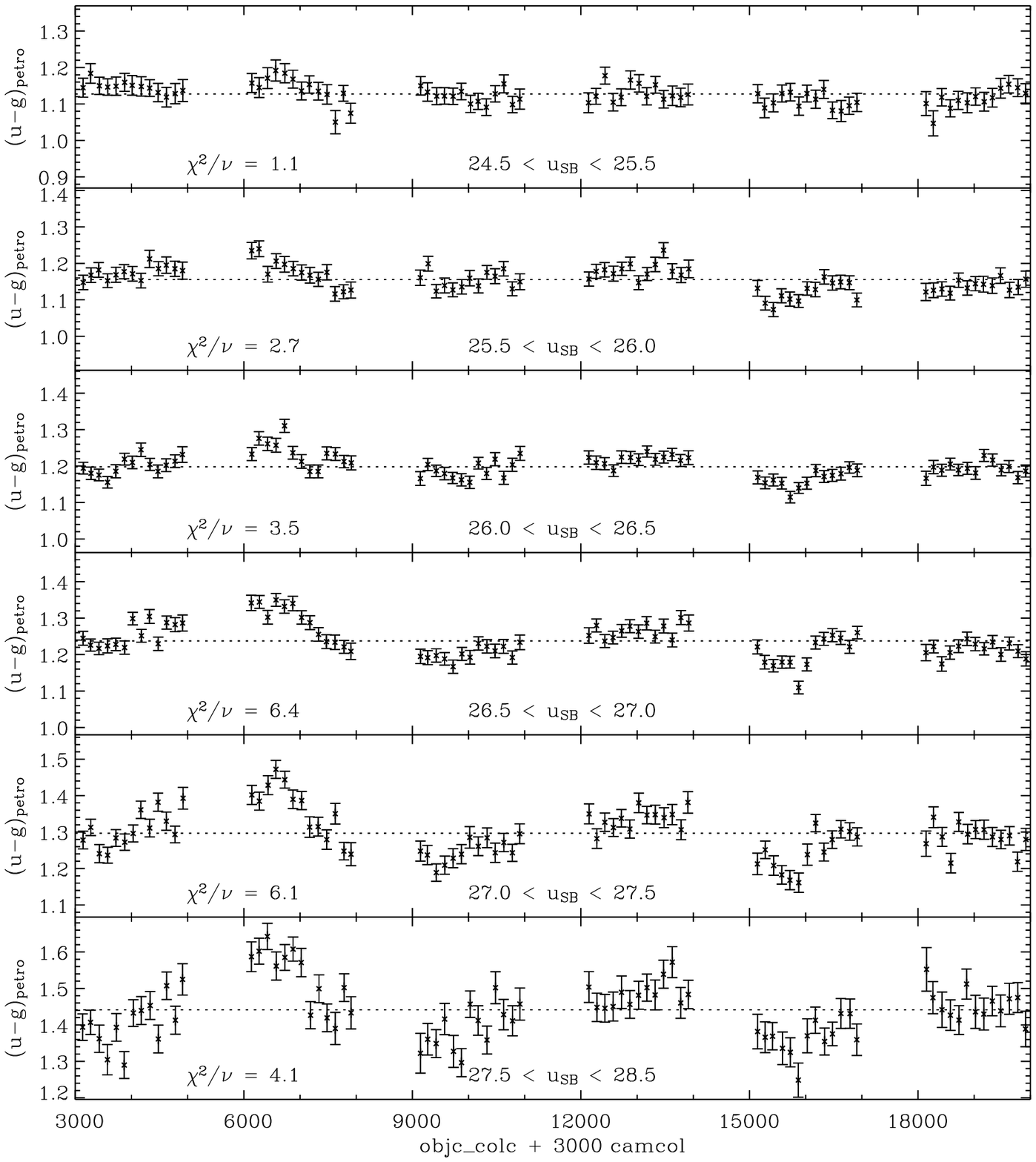}
\caption{Variation in the average observed $u-g$ galaxy colors 
  as a function of the pixel position (`objc\_colc') and the camera column
  (`camcol').  The panels represent different SB bins. The reduced $\chi^2$
  values on the assumption of no variation with camera position are shown in
  the panels, with the dotted lines showing the mean color. Note that the SDSS
  cameras operate in drift-scan mode so that any systematic error only depends
  on the detector column (and not on the detector row).}
\label{fig:u-camera-corr}
\end{figure*}

The variation of the average $u-g$ galaxy color, as a function of camera
column, is shown in Fig.~\ref{fig:u-camera-corr} for different SB bins.
There is a significant non-astrophysical variation of the average color and
the amplitude of the variation increases with decreasing SB.  The variation is
reduced by using model colors because the effective aperture is smaller.  The
peak-to-peak systematic variation of $u-g$ color is around 0.3--0.4 for
Petrosian magnitudes (lower two panels of Fig.~\ref{fig:u-camera-corr}) and
around 0.1--0.2 for model magnitudes, for galaxies with $u$-band surface
brightnesses between 27 and 28 mag arcsec$^{-2}$.

On the assumption that the dominate systematic error is due to sky-subtraction
errors, we determined linear offsets in SB as a function of pixel position
that minimized the color variation.  This was done by using the three lowest
SB bins and fitting a cubic polynomial to the implied offsets for each camera
column separately.  The $\upetro$ fluxes were redetermined and the process was
iterated until the results converged.  The mean flux offset was normalized to
zero.  The polynomial coefficients are given in Table~\ref{tab:u-correct} and
should be applied so that
\begin{equation}
  u'_{\rm flux} = u_{\rm flux} + P({\rm objc\_colc}) \, \pi \, (2 R_r)^2
\label{eqn:u-band-correct}
\end{equation}
where: $u_{\rm flux}$ is the pipeline Petrosian flux in units of `maggies'
[$-2.5\log({\rm maggies}) = {\rm mag}$]; $P$ is the polynomial function
of the pixel position, which is different for each `camcol'; and $R_r$ is the
Petrosian radius in the $r$-band. Note that the SDSS databases use asinh
magnitudes \citep{LGS99}, which can be converted to flux by
\begin{equation}
X_{\rm flux} = 
  \sinh \left[ -\frac{\ln(10)}{2.5} X_{\rm mag} - \ln(b) \right] 2 b
\label{eqn:asinh-convert}
\end{equation}
where the $b$ values are (1.4, 0.9, 1.2, 1.8, 7.4) $\times 10^{-10}$ for
$ugriz$, respectively \citep{stoughton02}.

\begin{table}
\caption{Polynomial coefficients for the $u$-band Petrosian flux
  corrections as a function of objc\_colc}
\label{tab:u-correct}
\begin{center}
\begin{tabular}{crrrr} \hline
camcol& $p_0\,/\,10^{-12}$& $p_1\,/\,10^{-15}$& 
$p_2\,/\,10^{-18}$& $p_3\,/\,10^{-21}$ \\ 
\\
 1& $ 0.187$& $-2.976$& $ 3.767$& $-1.101$\\ 
 2& $ 0.902$& $ 5.103$& $-6.910$& $ 2.063$\\ 
 3& $-0.616$& $-1.674$& $ 2.391$& $-0.749$\\ 
 4& $ 0.485$& $-1.027$& $ 1.460$& $-0.456$\\ 
 5& $ 0.481$& $-6.971$& $ 7.650$& $-2.181$\\ 
 6& $ 0.242$& $-1.933$& $ 2.514$& $-0.907$\\ 
\hline
\end{tabular}
\end{center}
\begin{flushleft}
See Eqs.~\ref{eqn:u-band-correct}--\ref{eqn:asinh-convert} for how to apply
the correction, and see Fig.~\ref{fig:u-correct} for an illustration of the
offsets for each of the camera columns. Note that these corrections were
determined for {\tt photo} v.~5.4 and a Southern Survey coadded catalog.
Similar offsets are observed using earlier versions of the pipeline and for
different regions of the sky.
\end{flushleft}
\end{table}

\begin{figure*}
\includegraphics[width=0.9\textwidth]{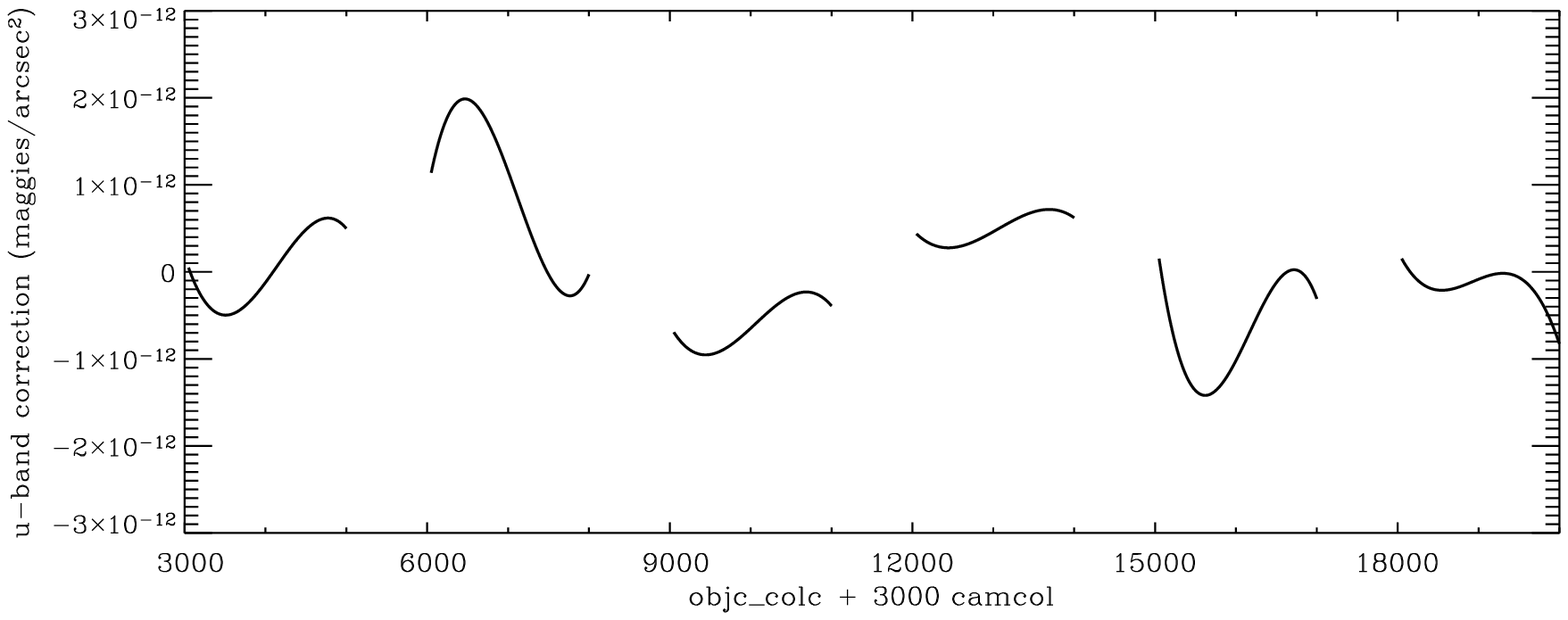}
\caption{Estimated flux offsets required to correct sky-subtraction
  errors in the $u$-band.  The lines represent the polynomial fits to the data
  given in Table~\ref{tab:u-correct}.  The units are linear in flux where
  10$^{-12}$ maggies corresponds to a magnitude of 30 [i.e.\ $-2.5\log({\rm
    maggies}) = {\rm mag}$]. The average of the flux offsets was
  normalized to zero. Note that these offsets were determined using a Southern
  Survey coadded catalog where there was some smoothing over objc\_colc
  (FWHM$\sim$50--200) because some drift-scan runs were offset slightly in
  DEC.}
\label{fig:u-correct}
\end{figure*}

\begin{figure*}
\includegraphics[width=0.9\textwidth]{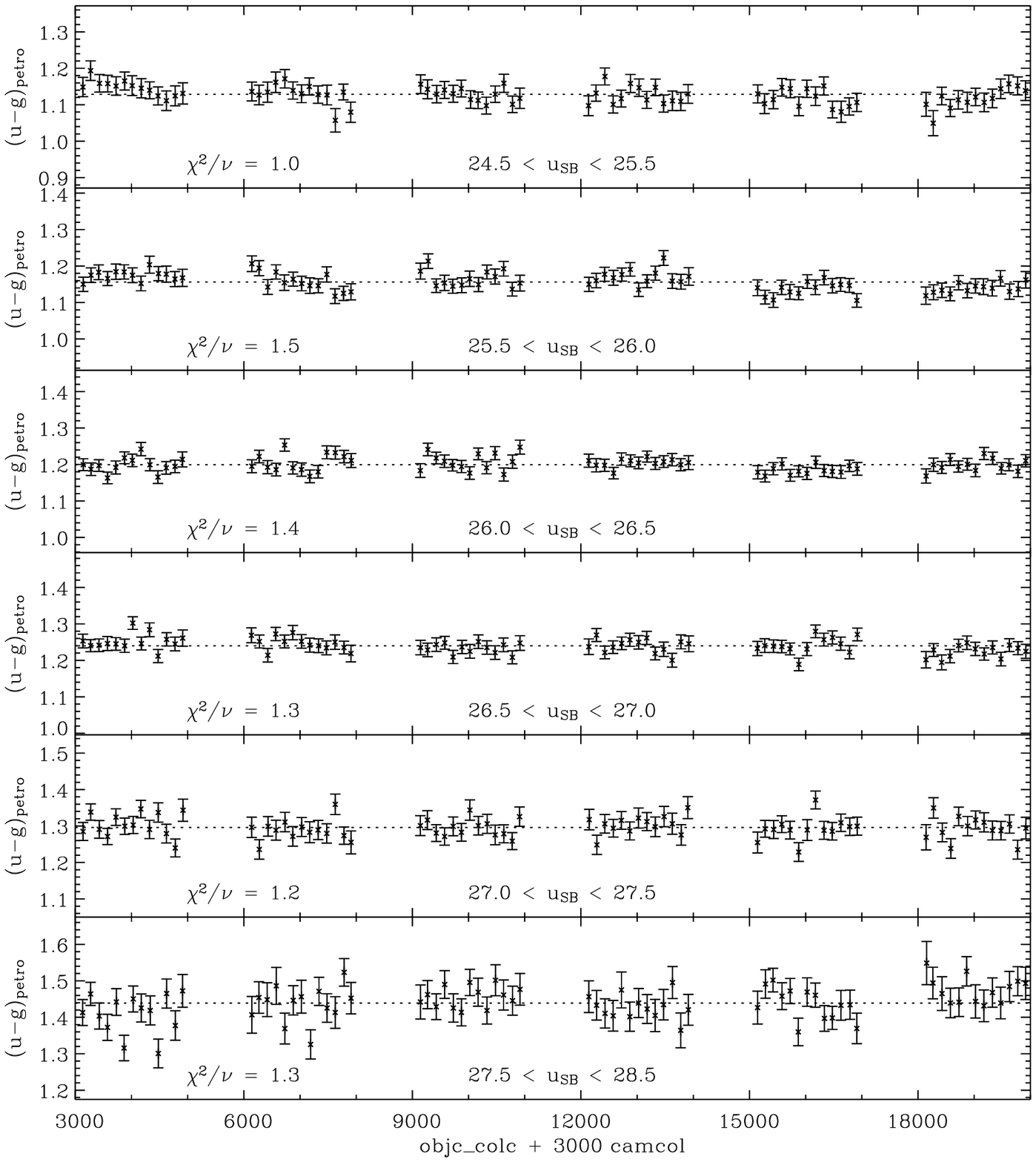}
\caption{Same as Fig.~\ref{fig:u-camera-corr} 
  except the $u$-band fluxes have been corrected using the functions given in
  Table~\ref{tab:u-correct}.}
\label{fig:u-camera-corr2}
\label{lastpage}
\end{figure*}

A plot of the offsets is shown in Fig.~\ref{fig:u-correct} and the variation
in the $u-g$ color after correction is shown in Fig.~\ref{fig:u-camera-corr2}.
After correction, there is no significant variation with camera column for
each SB bin (reduced $\chi^2\sim1$). This implies that our assumption that sky
subtraction was the dominate systematic error varying with camera column was
valid. For galaxies with low surface brightnesses (28 mag arcsec$^{-2}$ over 
the aperture), the correction can be up to $\pm0.3$ mag depending on
the pixel position.

\bibliographystyle{mn2e}
\bibliography{cosmology,fioc-rv,surveys,two-d-f,stars,galaxies,general}

\begin{thebibliography}{}

\bibitem[\protect\citeauthoryear{Abazajian et~al.,}{Abazajian
  et~al.}{2003}]{abazajian03}
Abazajian K.,  et~al., 2003, \aj, 126, 2081

\bibitem[\protect\citeauthoryear{Abazajian et~al.,}{Abazajian
  et~al.}{2004}]{abazajian04}
Abazajian K.,  et~al., 2004, \aj, 128, 502

\bibitem[\protect\citeauthoryear{{Baldry} \& Glazebrook}{{Baldry} \&
  Glazebrook}{2003}]{BG03}
{Baldry} I.~K.,  Glazebrook K.,  2003, \apj, 593, 258

\bibitem[\protect\citeauthoryear{{Blanton}, {Brinkmann} et~al.,}{{Blanton}
  et~al.}{2003a}]{blanton03kcorr}
{Blanton} M.~R.,  {Brinkmann} J.,    et~al., 2003a, \aj, 125, 2348

\bibitem[\protect\citeauthoryear{{Blanton}, {Hogg} et~al.,}{{Blanton}
  et~al.}{2003b}]{blanton03ld}
{Blanton} M.~R.,  {Hogg} D.~W.,    et~al., 2003b, \apj, 592, 819

\bibitem[\protect\citeauthoryear{{Blanton}, {Hogg} et~al.,}{{Blanton}
  et~al.}{2003c}]{blanton03broadband}
{Blanton} M.~R.,  {Hogg} D.~W.,    et~al., 2003c, \apj, 594, 186

\bibitem[\protect\citeauthoryear{{Blanton}, {Lin}, {Lupton}, {Maley}, {Young},
  {Zehavi} \& {Loveday}}{{Blanton} et~al.}{2003d}]{blanton03tile}
{Blanton} M.~R.,  {Lin} H.,  {Lupton} R.~H.,  {Maley} F.~M.,  {Young} N.,
  {Zehavi} I.,    {Loveday} J.,  2003d, \aj, 125, 2276

\bibitem[\protect\citeauthoryear{{Blanton}, {Lupton}, {Schlegel}, {Strauss},
  {Brinkmann}, {Fukugita} \& {Loveday}}{{Blanton} et~al.}{2004}]{blanton04}
{Blanton} M.~R.,  {Lupton} R.~H.,  {Schlegel} D.~J.,  {Strauss} M.~A.,
  {Brinkmann} J.,  {Fukugita} M.,    {Loveday} J.,  2004, \apj, submitted
  (astro-ph/0410164)

\bibitem[\protect\citeauthoryear{{Bruzual} \& {Charlot}}{{Bruzual} \&
  {Charlot}}{2003}]{BC03}
{Bruzual} G.,  {Charlot} S.,  2003, \mnras, 344, 1000

\bibitem[\protect\citeauthoryear{{Budav{\' a}ri}, Szalay, Charlot
  et~al.,}{{Budav{\' a}ri} et~al.}{2005}]{budavari05}
{Budav{\' a}ri} T.,  Szalay A.~S.,  Charlot S.,    et~al., 2005, \apj, in press
  (astro-ph/0411305)

\bibitem[\protect\citeauthoryear{{Christlein}, {McIntosh} \&
  {Zabludoff}}{{Christlein} et~al.}{2004}]{CMZ04}
{Christlein} D.,  {McIntosh} D.~H.,    {Zabludoff} A.~I.,  2004, \apj, 611, 795

\bibitem[\protect\citeauthoryear{{Colless}, {Dalton}, {Maddox}
  et~al.,}{{Colless} et~al.}{2001}]{colless01}
{Colless} M.,  {Dalton} G.,  {Maddox} S.,    et~al., 2001, \mnras, 328, 1039

\bibitem[\protect\citeauthoryear{{Connolly}, {Szalay}, {Dickinson}, {Subbarao}
  \& {Brunner}}{{Connolly} et~al.}{1997}]{connolly97}
{Connolly} A.~J.,  {Szalay} A.~S.,  {Dickinson} M.,  {Subbarao} M.~U.,
  {Brunner} R.~J.,  1997, \apjl, 486, L11

\bibitem[\protect\citeauthoryear{{Cowie}, {Songaila} \& {Barger}}{{Cowie}
  et~al.}{1999}]{CSB99}
{Cowie} L.~L.,  {Songaila} A.,    {Barger} A.~J.,  1999, \aj, 118, 603

\bibitem[\protect\citeauthoryear{{Cowie}, {Songaila}, {Hu} \& {Cohen}}{{Cowie}
  et~al.}{1996}]{cowie96}
{Cowie} L.~L.,  {Songaila} A.,  {Hu} E.~M.,    {Cohen} J.~G.,  1996, \aj, 112,
  839

\bibitem[\protect\citeauthoryear{{Cross}, {Driver}, {Couch} et~al.,}{{Cross}
  et~al.}{2001}]{cross01}
{Cross} N.,  {Driver} S.~P.,  {Couch} W.,    et~al., 2001, \mnras, 324, 825

\bibitem[\protect\citeauthoryear{{Davis}, {Faber}, {Newman} et~al.,}{{Davis}
  et~al.}{2003}]{davis03}
{Davis} M.,  {Faber} S.~M.,  {Newman} J.,    et~al., 2003, \procspie, 4834, 161

\bibitem[\protect\citeauthoryear{{de Vaucouleurs}}{{de
  Vaucouleurs}}{1959}]{deVaucouleurs59}
{de Vaucouleurs} G.,  1959, Handb. Physik, 53, 311

\bibitem[\protect\citeauthoryear{{Eisenstein}, {Annis}, {Gunn}
  et~al.,}{{Eisenstein} et~al.}{2001}]{eisenstein01}
{Eisenstein} D.~J.,  {Annis} J.,  {Gunn} J.~E.,    et~al., 2001, \aj, 122, 2267

\bibitem[\protect\citeauthoryear{{Fioc} \& {Rocca-Volmerange}}{{Fioc} \&
  {Rocca-Volmerange}}{1997}]{FR97}
{Fioc} M.,  {Rocca-Volmerange} B.,  1997, \aap, 326, 950

\bibitem[\protect\citeauthoryear{{Fioc} \& {Rocca-Volmerange}}{{Fioc} \&
  {Rocca-Volmerange}}{1999}]{FR99}
{Fioc} M.,  {Rocca-Volmerange} B.,  1999, preprint (astro-ph/9912179)

\bibitem[\protect\citeauthoryear{{Freeman}}{{Freeman}}{1970}]{Freeman70}
{Freeman} K.~C.,  1970, \apj, 160, 811

\bibitem[\protect\citeauthoryear{{Fukugita}, {Ichikawa}, {Gunn}, {Doi},
  {Shimasaku} \& {Schneider}}{{Fukugita} et~al.}{1996}]{fukugita96}
{Fukugita} M.,  {Ichikawa} T.,  {Gunn} J.~E.,  {Doi} M.,  {Shimasaku} K.,
  {Schneider} D.~P.,  1996, \aj, 111, 1748

\bibitem[\protect\citeauthoryear{{Gunn}, {Carr}, {Rockosi} et~al.,}{{Gunn}
  et~al.}{1998}]{gunn98}
{Gunn} J.~E.,  {Carr} M.,  {Rockosi} C.,    et~al., 1998, \aj, 116, 3040

\bibitem[\protect\citeauthoryear{{Hogg}, Baldry, Blanton \& Eisenstein}{{Hogg}
  et~al.}{2002}]{hogg02kcorr}
{Hogg} D.~W.,  Baldry I.~K.,  Blanton M.~R.,    Eisenstein D.~J.,  2002,
  preprint (astro-ph/0210394)

\bibitem[\protect\citeauthoryear{{Hogg}, {Finkbeiner}, {Schlegel} \&
  {Gunn}}{{Hogg} et~al.}{2001}]{hogg01}
{Hogg} D.~W.,  {Finkbeiner} D.~P.,  {Schlegel} D.~J.,    {Gunn} J.~E.,  2001,
  \aj, 122, 2129

\bibitem[\protect\citeauthoryear{{Hopkins}}{{Hopkins}}{2004}]{Hopkins04}
{Hopkins} A.~M.,  2004, \apj, 615, 209

\bibitem[\protect\citeauthoryear{{Hopkins}, {Connolly}, {Haarsma} \&
  {Cram}}{{Hopkins} et~al.}{2001}]{hopkins01}
{Hopkins} A.~M.,  {Connolly} A.~J.,  {Haarsma} D.~B.,    {Cram} L.~E.,  2001,
  \aj, 122, 288

\bibitem[\protect\citeauthoryear{{Hopkins}, {Miller}, {Nichol}
  et~al.,}{{Hopkins} et~al.}{2003}]{hopkins03}
{Hopkins} A.~M.,  {Miller} C.~J.,  {Nichol} R.~C.,    et~al., 2003, \apj, 599,
  971

\bibitem[\protect\citeauthoryear{{Ilbert}, {Tresse}, {Zucca} et~al.,}{{Ilbert}
  et~al.}{2004}]{ilbert04}
{Ilbert} O.,  {Tresse} L.,  {Zucca} E.,    et~al., 2004, \aap, submitted
  (astro-ph/0409134)

\bibitem[\protect\citeauthoryear{{Jones}, {Saunders}, {Colless}
  et~al.,}{{Jones} et~al.}{2004}]{jones04}
{Jones} D.~H.,  {Saunders} W.,  {Colless} M.,    et~al., 2004, \mnras, 355, 747

\bibitem[\protect\citeauthoryear{{Le F{\` e}vre}, {Vettolani}, {Garilli}
  et~al.,}{{Le F{\` e}vre} et~al.}{2004}]{lefevre04}
{Le F{\` e}vre} O.,  {Vettolani} G.,  {Garilli} B.,    et~al., 2004, \aap,
  submitted (astro-ph/0409133)

\bibitem[\protect\citeauthoryear{{Lilly}, {Le Fevre}, {Hammer} \&
  {Crampton}}{{Lilly} et~al.}{1996}]{lilly96}
{Lilly} S.~J.,  {Le Fevre} O.,  {Hammer} F.,    {Crampton} D.,  1996, \apjl,
  460, L1

\bibitem[\protect\citeauthoryear{{Lin}, {Yee}, {Carlberg}, {Morris}, {Sawicki},
  {Patton}, {Wirth} \& {Shepherd}}{{Lin} et~al.}{1999}]{lin99}
{Lin} H.,  {Yee} H.~K.~C.,  {Carlberg} R.~G.,  {Morris} S.~L.,  {Sawicki} M.,
  {Patton} D.~R.,  {Wirth} G.,    {Shepherd} C.~W.,  1999, \apj, 518, 533

\bibitem[\protect\citeauthoryear{{Lupton}, {Gunn}, {Ivezi{\' c}}, {Knapp},
  {Kent} \& {Yasuda}}{{Lupton} et~al.}{2001}]{lupton01}
{Lupton} R.~H.,  {Gunn} J.~E.,  {Ivezi{\' c}} Z.,  {Knapp} G.~R.,  {Kent} S.,
   {Yasuda} N.,  2001, in Harnden F.~R.,  Primini F.~A.,   Payne H.~E.,  eds,
  ASP Conf. Ser., Vol. 238, Astron. Data Analysis Software Syst. X. ASP, San
  Francisco, p.~269

\bibitem[\protect\citeauthoryear{{Lupton}, {Gunn} \& {Szalay}}{{Lupton}
  et~al.}{1999}]{LGS99}
{Lupton} R.~H.,  {Gunn} J.~E.,    {Szalay} A.~S.,  1999, \aj, 118, 1406

\bibitem[\protect\citeauthoryear{{Madau}, {Ferguson}, {Dickinson},
  {Giavalisco}, {Steidel} \& {Fruchter}}{{Madau} et~al.}{1996}]{madau96}
{Madau} P.,  {Ferguson} H.~C.,  {Dickinson} M.~E.,  {Giavalisco} M.,  {Steidel}
  C.~C.,    {Fruchter} A.,  1996, \mnras, 283, 1388

\bibitem[\protect\citeauthoryear{{Madau}, {Pozzetti} \& {Dickinson}}{{Madau}
  et~al.}{1998}]{MPD98}
{Madau} P.,  {Pozzetti} L.,    {Dickinson} M.,  1998, \apj, 498, 106

\bibitem[\protect\citeauthoryear{{Martin} et~al.,}{{Martin}
  et~al.}{2003}]{martin03}
{Martin} C.,  et~al., 2003, \procspie, 4854, 336

\bibitem[\protect\citeauthoryear{{Martin}, {Fanson}, {Schiminovich}
  et~al.,}{{Martin} et~al.}{2005}]{martin05}
{Martin} D.~C.,  {Fanson} J.,  {Schiminovich} D.,    et~al., 2005, \apj, in
  press (astro-ph/0411302)

\bibitem[\protect\citeauthoryear{{Milliard}, {Donas}, {Laget}, {Armand} \&
  {Vuillemin}}{{Milliard} et~al.}{1992}]{milliard92}
{Milliard} B.,  {Donas} J.,  {Laget} M.,  {Armand} C.,    {Vuillemin} A.,
  1992, \aap, 257, 24

\bibitem[\protect\citeauthoryear{Oke \& {Gunn}}{Oke \& {Gunn}}{1983}]{OG83}
Oke J.~B.,  {Gunn} J.~E.,  1983, ApJ, 266, 713

\bibitem[\protect\citeauthoryear{{Percival} et~al.,}{{Percival}
  et~al.}{2001}]{percival01}
{Percival} W.~J.,  et~al., 2001, \mnras, 327, 1297

\bibitem[\protect\citeauthoryear{{Petrosian}}{{Petrosian}}{1976}]{Petrosian76}
{Petrosian} V.,  1976, \apjl, 209, L1

\bibitem[\protect\citeauthoryear{{Pier}, {Munn}, {Hindsley}, {Hennessy},
  {Kent}, {Lupton} \& {Ivezi{\' c}}}{{Pier} et~al.}{2003}]{pier03}
{Pier} J.~R.,  {Munn} J.~A.,  {Hindsley} R.~B.,  {Hennessy} G.~S.,  {Kent}
  S.~M.,  {Lupton} R.~H.,    {Ivezi{\' c}} {\v Z}.,  2003, \aj, 125, 1559

\bibitem[\protect\citeauthoryear{{Richards}, {Fan}, {Newberg}
  et~al.,}{{Richards} et~al.}{2002}]{richards02}
{Richards} G.~T.,  {Fan} X.,  {Newberg} H.~J.,    et~al., 2002, \aj, 123, 2945

\bibitem[\protect\citeauthoryear{{Schechter}}{{Schechter}}{1976}]{Schechter76}
{Schechter} P.,  1976, \apj, 203, 297

\bibitem[\protect\citeauthoryear{{Schiminovich}, {Ilbert}, {Arnouts}
  et~al.,}{{Schiminovich} et~al.}{2005}]{schiminovich05}
{Schiminovich} D.,  {Ilbert} O.,  {Arnouts} S.,    et~al., 2005, \apj, in press
  (astro-ph/0411424)

\bibitem[\protect\citeauthoryear{{Schlegel}, {Finkbeiner} \&
  {Davis}}{{Schlegel} et~al.}{1998}]{SFD98}
{Schlegel} D.~J.,  {Finkbeiner} D.~P.,    {Davis} M.,  1998, \apj, 500, 525

\bibitem[\protect\citeauthoryear{{Schmidt}}{{Schmidt}}{1968}]{Schmidt68}
{Schmidt} M.,  1968, \apj, 151, 393

\bibitem[\protect\citeauthoryear{{Smith}, {Tucker}, {Kent} et~al.,}{{Smith}
  et~al.}{2002}]{smith02}
{Smith} J.~A.,  {Tucker} D.~L.,  {Kent} S.,    et~al., 2002, \aj, 123, 2121

\bibitem[\protect\citeauthoryear{{Somerville}, {Lee}, {Ferguson}, {Gardner},
  {Moustakas} \& {Giavalisco}}{{Somerville} et~al.}{2004}]{somerville04cv}
{Somerville} R.~S.,  {Lee} K.,  {Ferguson} H.~C.,  {Gardner} J.~P.,
  {Moustakas} L.~A.,    {Giavalisco} M.,  2004, \apjl, 600, L171

\bibitem[\protect\citeauthoryear{{Steidel}, {Adelberger}, {Giavalisco},
  {Dickinson} \& {Pettini}}{{Steidel} et~al.}{1999}]{steidel99}
{Steidel} C.~C.,  {Adelberger} K.~L.,  {Giavalisco} M.,  {Dickinson} M.,
  {Pettini} M.,  1999, \apj, 519, 1

\bibitem[\protect\citeauthoryear{{Stoughton}, {Lupton} et~al.,}{{Stoughton}
  et~al.}{2002}]{stoughton02}
{Stoughton} C.,  {Lupton} R.~H.,    et~al., 2002, \aj, 123, 485

\bibitem[\protect\citeauthoryear{{Strateva}, {Ivezi{\' c}}, {Knapp}
  et~al.,}{{Strateva} et~al.}{2001}]{strateva01}
{Strateva} I.,  {Ivezi{\' c}} {\v Z}.,  {Knapp} G.~R.,    et~al., 2001, \aj,
  122, 1861

\bibitem[\protect\citeauthoryear{{Strauss}, {Weinberg}, {Lupton}
  et~al.,}{{Strauss} et~al.}{2002}]{strauss02}
{Strauss} M.~A.,  {Weinberg} D.~H.,  {Lupton} R.~H.,    et~al., 2002, \aj, 124,
  1810

\bibitem[\protect\citeauthoryear{{Sullivan}, {Treyer}, {Ellis}, {Bridges},
  {Milliard} \& {Donas}}{{Sullivan} et~al.}{2000}]{sullivan00}
{Sullivan} M.,  {Treyer} M.~A.,  {Ellis} R.~S.,  {Bridges} T.~J.,  {Milliard}
  B.,    {Donas} J.~.,  2000, \mnras, 312, 442

\bibitem[\protect\citeauthoryear{{Treyer}, {Ellis}, {Milliard}, {Donas} \&
  {Bridges}}{{Treyer} et~al.}{1998}]{treyer98}
{Treyer} M.~A.,  {Ellis} R.~S.,  {Milliard} B.,  {Donas} J.,    {Bridges}
  T.~J.,  1998, \mnras, 300, 303

\bibitem[\protect\citeauthoryear{{Uomoto}, {Smee}, {Rockosi} et~al.,}{{Uomoto}
  et~al.}{1999}]{uomoto99}
{Uomoto} A.,  {Smee} S.,  {Rockosi} C.,    et~al., 1999, Bull.\ American
  Astron.\ Soc., 31, 1501

\bibitem[\protect\citeauthoryear{{Vanden Berk}, Schneider, Richards
  et~al.,}{{Vanden Berk} et~al.}{2005}]{vandenberk05}
{Vanden Berk} D.,  Schneider D.,  Richards G.,    et~al., 2005, \aj, in press

\bibitem[\protect\citeauthoryear{{Wilson}, {Cowie}, {Barger} \&
  {Burke}}{{Wilson} et~al.}{2002}]{wilson02}
{Wilson} G.,  {Cowie} L.~L.,  {Barger} A.~J.,    {Burke} D.~J.,  2002, \aj,
  124, 1258

\bibitem[\protect\citeauthoryear{{Wolf}, {Meisenheimer}, {Rix}, {Borch}, {Dye}
  \& {Kleinheinrich}}{{Wolf} et~al.}{2003}]{wolf03}
{Wolf} C.,  {Meisenheimer} K.,  {Rix} H.-W.,  {Borch} A.,  {Dye} S.,
  {Kleinheinrich} M.,  2003, \aap, 401, 73

\bibitem[\protect\citeauthoryear{Wyder, Treyer, Milliard et~al.,}{Wyder
  et~al.}{2005}]{wyder05}
Wyder T.~K.,  Treyer M.~A.,  Milliard B.,    et~al., 2005, \apj, in press
  (astro-ph/0411364)

\bibitem[\protect\citeauthoryear{{York} et~al.,}{{York}  et~al.}{2000}]{york00}
{York} D.~G.,  et~al., 2000, \aj, 120, 1579

\end{thebibliography}

\end{document}